\documentclass[aps,pre,citeautoscript,twocolumn,showpacs,floatfix,superscriptaddress,graphics,float]{revtex4-1}

\usepackage{amsmath,amssymb,graphicx,color,braket,hyphenat,makeidx,xcolor,dsfont,subfigure}
\usepackage[ocgcolorlinks,colorlinks=true,linkcolor=blue,citecolor=red,linktocpage=true]{hyperref}

\usepackage[english]{babel}
\newcommand{\blue}{\color{\blue}}

\newcommand{\tr}{\mathrm{Tr}}

\begin{document}

\title{Optimizing autonomous thermal machines powered by energetic coherence}

\author{Kenza Hammam}
\affiliation{Equipe des Sciences de la Mati\`ere et du Rayonnement (ESMaR), Facult\'e des Sciences, Universit\'e Mohammed V, Av. Ibn Battouta, B.P. 1014, Agdal, Rabat, Morocco.}\affiliation{International Center for Theoretical Physics ICTP, Strada Costiera 11, I-34151, Trieste, Italy}

\author{Yassine Hassouni}
\affiliation{Equipe des Sciences de la Mati\`ere et du Rayonnement (ESMaR), Facult\'e des Sciences, Universit\'e Mohammed V, Av. Ibn Battouta, B.P. 1014, Agdal, Rabat, Morocco.}

\author{Rosario Fazio}
\affiliation{International Center for Theoretical Physics ICTP, Strada Costiera 11, I-34151, Trieste, Italy}
\affiliation{Dipartimento di Fisica, Universit\`a di Napoli "Federico II", Monte S. Angelo, I-80126 Napoli, Italy}

\author{Gonzalo Manzano}
\affiliation{Institute for Quantum Optics and Quantum Information (IQOQI), Austrian Academy of Sciences, Boltzmanngasse 3, 1090 Vienna, Austria}
\affiliation{International Center for Theoretical Physics ICTP, Strada Costiera 11, I-34151, Trieste, Italy}

\begin{abstract}
The characterization and control of quantum effects in the performance of thermodynamic tasks may open new avenues for small thermal machines working in the nanoscale. We study the impact of coherence in the energy basis in the operation of a small thermal machine which can act either as a heat engine or as a refrigerator. We show that input coherence may enhance the machine performance and allow it to operate in otherwise forbidden regimes. Moreover, our results also indicate that, in some cases, coherence may also be detrimental, rendering optimization of particular models a crucial task for benefiting from coherence-induced enhancements.  
\end{abstract}

%\pacs{
%05.70.Ln,  %    Nonequilibrium and irreversible thermodynamics
%05.70.-a   %    Thermodynamics
%05.30.-d   %    Quantum statistical mechanics
%03.67.-a   %    Quantum information
%} 
\maketitle

\section{Introduction}

Thermodynamics was originally conceived in the XIX century to describe and eventually improve steam engines~\cite{Carnot,Prigogine}. It has proven to be one of the most successful theories in physics, leading to a general understanding of the physical properties of macroscopic systems, with multidisciplinary applications. Recent decades have seen a growth of interest in the thermodynamics of microscopic systems~\cite{QTD1,QTD2}. Progress in experimental techniques to manipulate quantum systems in the laboratory led us to even envision a future in which thermal devices may exploit genuine quantum effects, namely, quantum coherence~\cite{Scully:2003,Brandner:2015,Korzekwa:2016,Kurizki:2016} and quantum correlations~\cite{Acin:2015,Bera:2017,Zambrini:2018}. These effects intrinsically set quantum thermodynamics apart from its classical counterpart. Therefore clarifying their operational role and quantifying its impact becomes a crucial challenge. 
Pursuing this goal, many contributions have already considered the role of noise-induced (or degenerate) coherence~\cite{Scully:2011,Harbola:2012,Mukamel:2012,Dorfman:2013,Mitchison:2015,Brask:2015,GK:2015,Turkpence:2016,Cao:2018,Latune:2019}, non-thermal properties such as squeezing~\cite{Huang:2012,Rossnagel:2014,Correa:2014,Manzano:2016,Manzano:2020} or many-body correlations~\cite{Lutz:2009,Park:2013,Brunner:2014,Leggio:2016} on assisting thermodynamic tasks such as cooling and extracting work.

Since the introduction of the first operating prototype of a quantum heat engine by Scovil and Schultz-Dubois~\cite{Scovil:1959}, many models of quantum thermal machines have been put forward~\cite{Kosloff:2014,TQR:2018}. For example, quantum analogues of classical Carnot and Otto engines have been investigated~\cite{Quan:2007,Kosloff:2017}, paving the way for first experimental implementations of quantum heat engines in the last years~\cite{Rossnagel:2016,Serra:2019,Poem:2019,Poschinger:2019}.
A particularly appealing class of thermal machines are autonomous ones~\cite{TQR:2018}. They consist on self-contained models that avoid time-dependent Hamiltonians, since the later may implicitly require exta sources of coherence not explicitly accounted for~\cite{Aberg:2014,Saito:2020}. This makes autonomous machines an ideal platform to quantify the impact of coherence on the performance of thermodynamic tasks. Different models of autonomous thermal machines using few quantum levels, qubits or harmonic oscillators have been considered~\cite{Palao:2001,Linden:2010,Brunner:2012,Levy:2012}. Proposals for their implementation include different platforms, like quantum dots~\cite{Giovannetti:2013,Erdman:2018}, circuit QED architectures~\cite{Hofer:2016} or atoms in optical cavities~\cite{Mitchison:2016}. An autonomous quantum absorption refrigerator has been recently realized in the laboratory with trapped ions~\cite{Maslennikov:2019}.

\begin{figure}[t]
\includegraphics[width=1.0 \linewidth]{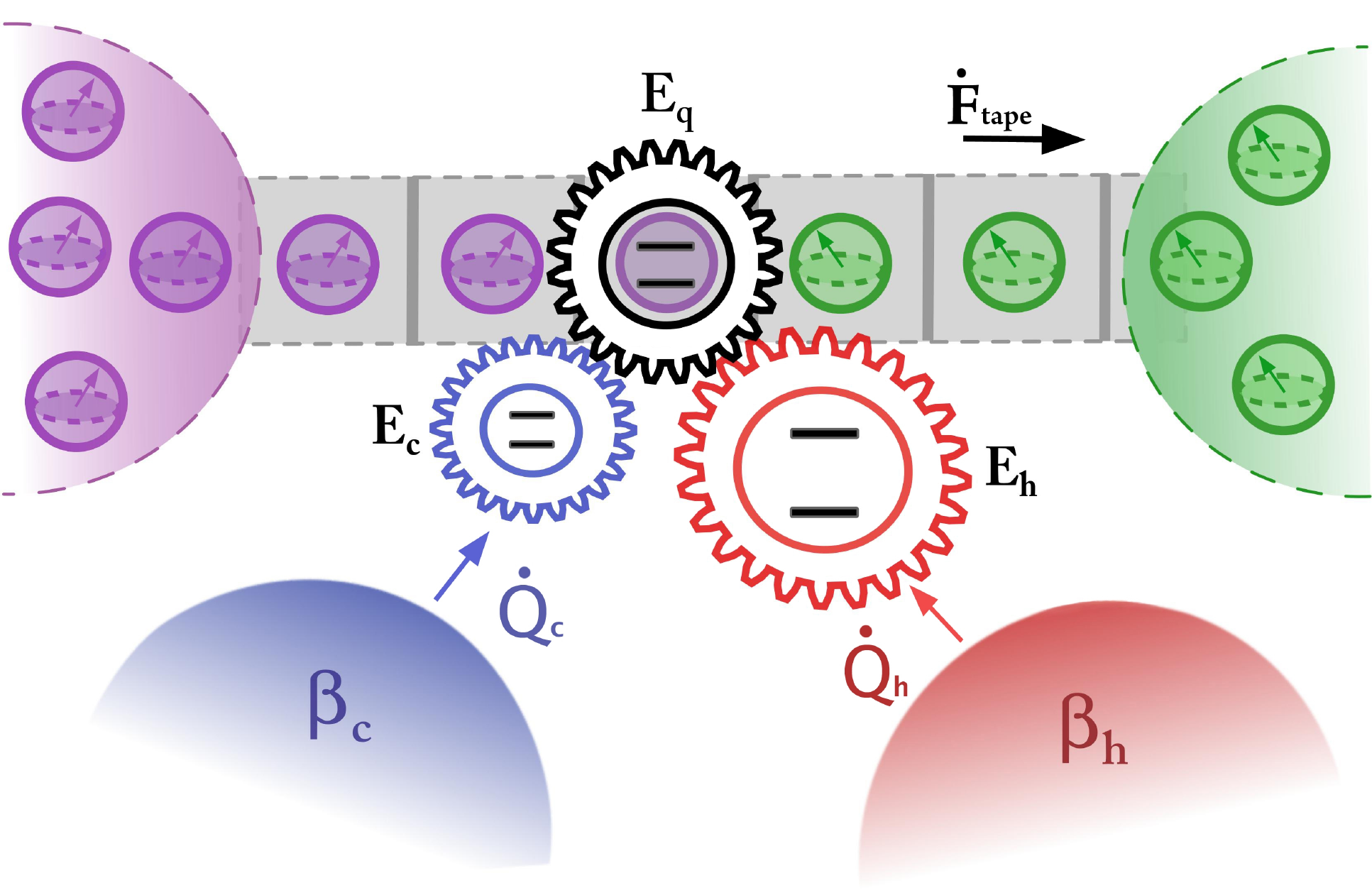}
\caption{Coherent thermal machine model. An ensemble of qubits prepared in an arbitrary state $\rho_\mathrm{q}$ (left purple area) is put on a tape moving from left to right to interact, once at a time, with the two machine qubits (center gears), after which they are collected (green right area). The setup allows either free energy extraction on the output qubits in the tape, powered by a heat current from hot to cold local baths (bottom red and blue areas), or refrigeration of the cold bath by consuming free energy of the incoming tape qubits.
}
\label{fig:1}
\end{figure}

Our aim here is to quantify and optimize the impact of energetic coherence on the performance of an autonomous thermal machine. By energetic coherence we refer to the coherence between non-degenerate energy levels, also called asymmetry under time-translations~\cite{Bartlett:2007,Streltsov:2017}. Energetic coherence is a resource of particular relevance in quantum thermodynamics~\cite{Woods:2018,Lostaglio:2019,Marvian:2019,Marvian:2020}. It constitutes an extra (independent) source of free energy~\cite{Janzing:2006,Kammerlander:2016}, allowing state transformations that would be otherwise impossible~\cite{Marvian:2014,Cwiklinski:2015,Lostaglio:2015,Uzdin:2016}, and has direct consequences on the entropy balance leading to the second law~\cite{Santos:2019,Plastina:2019}.

The model considered here has been recently introduced in Ref.~\cite{Manzano:2019}, although employed for different purposes. We consider one of the most prototypical quantum thermal machines, namely, a pair of two-level systems (or qubits) coupled to thermal baths at different temperatures~\cite{Brunner:2012,Kosloff:2017}. The machine works through the interaction with a stream of qubits (represented in a tape) via a sequence of energy-preserving collisions (see Fig.~\ref{fig:1}). 
Employing a collisional framework~\cite{Scarani:2002,Ziman:2005,Lorenzo:2017,Landi:2019} we relate heat an entropy fluxes from the baths with the transformations in the tape qubits to quantitatively address the performance of the machine. The presence of initial coherence in the tape qubits has a strong impact on the performance of the machine, that admits coherence-enhanced situations when optimization over the machine parameters is applied.

This paper has the following structure: In Sec.~\ref{sec:model} we describe the model and the dynamics for the thermal machine. The operation of the machine in steady-state conditions and its thermodynamic properties are explored in Sec.~\ref{sec:steady}. In Sec.~\ref{sec:regimes} we provide a description of the different modes of operation of the machine, highlighting the role of initial coherence in the tape  qubits. We then proceed to the optimization of the power of the machine in different regimes in Sec.~\ref{sec:optimization}, discussing the impact of coherence in both the power and the efficiency of the machine. Finally in Sec.~\ref{sec:conclusions}, we present our main conclusion and some perspectives for further research.

\section{Coherent machine model} \label{sec:model}

\begin{figure}[t]
\includegraphics[width=0.8 \linewidth]{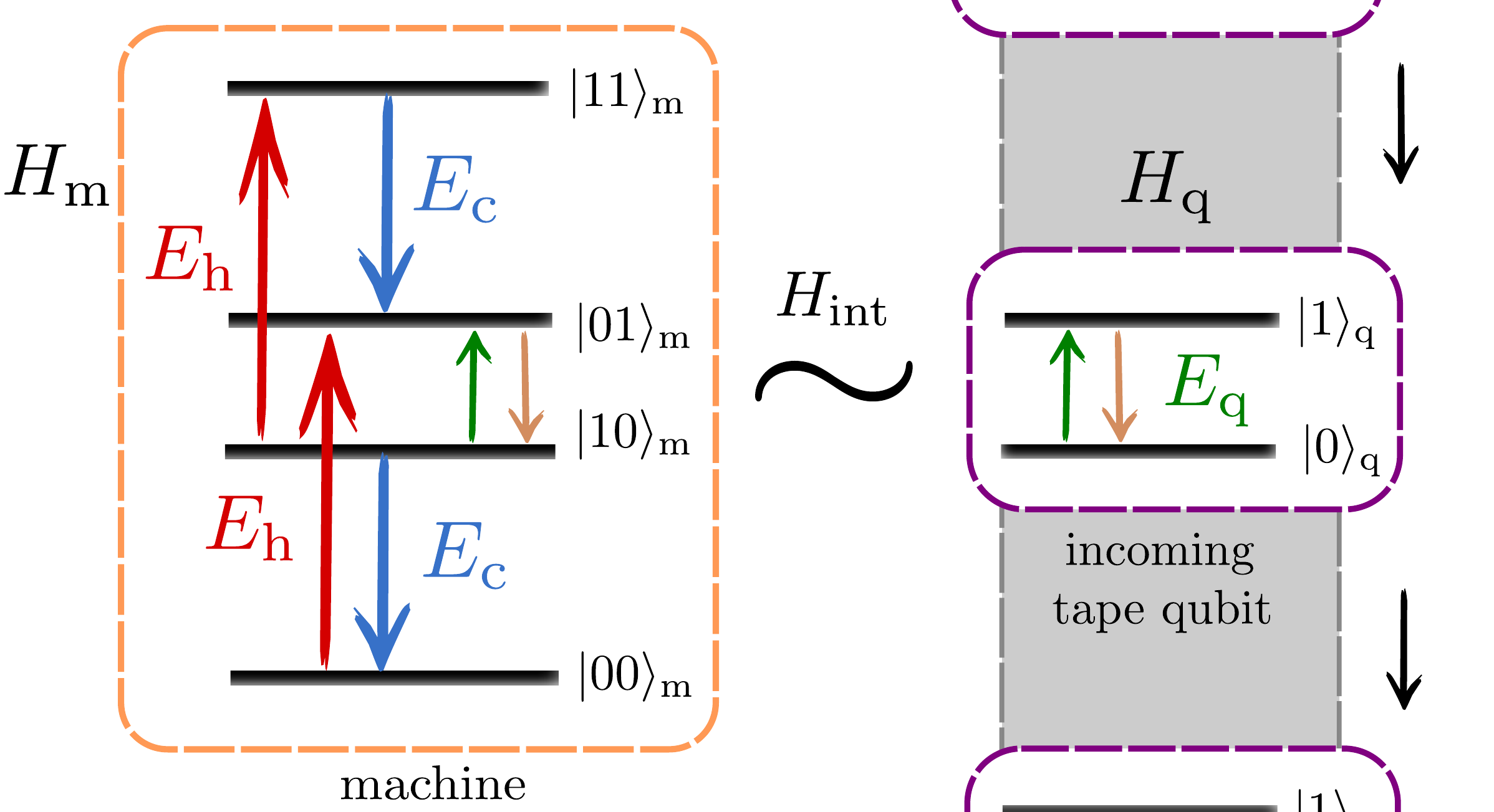}
\caption{Representation of the machine as a 4-level system interacting with an incoming qubit of the tape. The interaction $H_\mathrm{\rm int}$ with the resonance condition in Eq.~\eqref{eq:resonance}, allows the coherent exchange of excitations between the two inner machine levels and the tape qubit (green and orange arrows). The local reservoirs induce incoherent transitions on the machine levels with energy spacing $E_\mathrm{c}$ and $E_\mathrm{h}$ as represented, respectively, by the blue and red arrows.} 
\label{fig:2}
\end{figure}

The setup we employ along this study follows the configuration introduced in Ref.~\cite{Manzano:2019} and is sketched in Figs.~\ref{fig:1} and \ref{fig:2}. We consider a quantum thermal machine that consist of two qubits with energy spacing, $E_\mathrm{c}$ and $E_\mathrm{h}$ respectively, and Hamiltonian $H_\mathrm{m}= E_{\mathrm{c}} \ket{1}\bra{1}_\mathrm{c} \otimes \mathds{1}_\mathrm{h} + E_{\mathrm{h}}~\mathds{1}_{\mathrm{c}} \otimes \ket{1}\bra{1}_\mathrm{h}$, where we assume for concreteness $E_\mathrm{h} \geq E_\mathrm{c}$. The two machine qubits are weakly coupled to thermal reservoirs (or baths) with different inverse temperatures ($\beta_\mathrm{c} \geq \beta_\mathrm{h}$) acting only locally on each qubit. We refer to the machine qubits and baths as the ``cold'' and ``hot'' qubits or baths, respectively.

In addition, the machine sequentially interacts with a stream of qubits in a tape with fixed energy spacing 
\begin{equation} \label{eq:resonance}
E_\mathrm{q} = E_\mathrm{h} - E_\mathrm{c} ,
\end{equation}
and Hamiltonian $H_\mathrm{q} = E_\mathrm{q} \ket{1}\bra{1}_\mathrm{q}$, where the subscript $q$ stands for ``qubits''. All the qubits in the tape are assumed to start in the same generic initial state $\rho_\mathrm{q}$, and interact with the machine, one by one, trough a sequence of energy-preserving unitary maps of the form $U = e^{-i H_\mathrm{int} \tau}$ (in interaction picture), where $H_\mathrm{int}$ is the following three-body Hamiltonian:
\begin{equation} \label{eq:int}
 H_\mathrm{int} = g~(\sigma_\mathrm{c}^{+} \sigma_\mathrm{h}^{-} \sigma_\mathrm{q}^{+} + \sigma_\mathrm{c}^{-} \sigma_\mathrm{h}^{+} \sigma_\mathrm{q}^{-}), 
\end{equation}
where $\sigma_i^{-} \equiv \ket{0}\bra{1}_i$ denote the lowering operator of qubit $i= \mathrm{c},\mathrm{h},\mathrm{q}$ and $\sigma_i^{+} = (\sigma_i^{-})^\dagger$. We remark that strict energy conservation is enforced by the resonance condition~\eqref{eq:resonance}, which implies $[H_\mathrm{int}, H_\mathrm{m} + H_\mathrm{q}] =0$ and hence $[U, H_\mathrm{m} + H_\mathrm{q}]=0$. Here it is also useful to define $\phi= g \tau$ as the effective  strength of the collision, where $\tau$ denotes the interaction time. We are interested in the regime $\phi \ll 1$, corresponding to weak and fast collisions. 

In the absence of the qubit tape, the two qubits of the machine do not interact between them at all. In that case the contact with local thermal baths would produce the independent thermalization of each machine qubit to the cold and hot temperatures, $\beta_\mathrm{c}$ and $\beta_\mathrm{h}$, respectively. Even if in a product of thermal states, it is nonetheless interesting to look at the machine as composed of four levels (see Fig~\ref{fig:2}). Indeed the independent thermalization, would result in an effective population of the two inner levels $\{ \ket{10}_{\mathrm{m}}, \ket{01}_{\mathrm{m}} \}$~\cite{note:levels} (also called virtual qubit), which we denote as $p_{10}^{\mathrm{eq}}$ and $p_{01}^{\mathrm{eq}}$, showing a virtual inverse temperature $\beta_\mathrm{v}$ defined by the Gibbs ratio $\beta_{\mathrm{v}} E_\mathrm{q} \equiv \ln(p_{10}^\mathrm{eq}/p_{01}^{\mathrm{eq}})$~\cite{Brunner:2012}. It reads~ 
\begin{equation} \label{eq:virtual}
\beta_\mathrm{v} = \frac{\beta_\mathrm{h} E_\mathrm{h} - \beta_\mathrm{c} E_\mathrm{c}}{E_\mathrm{h} - E_\mathrm{c}}. 
\end{equation}
Depending on the values of $E_\mathrm{c}$ and $E_\mathrm{h}$, the virtual temperature can be tuned to take different values, making the machine suitable for the performance of thermodynamic tasks, such as power generation or refrigeration, by connecting a external system to the inner transition $\ket{10}_\mathrm{m} \leftrightarrow \ket{01}_\mathrm{m}$~\cite{Silva:2016,Huber:2017}. In particular, for the choice $E_\mathrm{h} \geq E_\mathrm{c}$ and $\beta_\mathrm{c} \geq \beta_\mathrm{h}$, we always have $\beta_\mathrm{v} \leq \beta_\mathrm{c}$, and obtain population inversion, $\beta_\mathrm{v} \leq 0$, if and only if $E_\mathrm{h}/E_\mathrm{c} \leq \beta_\mathrm{c}/\beta_\mathrm{h}$. As we will see in the following sections, the virtual temperature $\beta_\mathrm{v}$ plays an important role in the characterization of the operational regimes of the autonomous thermal machine considered here.

When introducing the interactions with the tape, the two qubits of the machine are allowed to exchange heat between them and with the qubits in the tape, evolving through a nonequilibrium stationary state. We assume that collisions occur at random times following Poissonian statistics with rate $r$. A monotonic evolution towards stationarity is enforced by the Markovian character of the collisional model employed here, where every collision uses a ``fresh'' qubit in the tape in the same state: 
\begin{equation} \label{eq:rho}
\rho_\mathrm{q} = 
 \begin{pmatrix}
p_0 & c \\
c^\ast & p_1
\end{pmatrix},
\end{equation}
with $p_0 + p_1 =1$ and $|c| \leq \sqrt{p_0 p_1}$ ensuring a positive semi-definite density operator. 

From the collisional model introduced here, we build a master equation in the GKLS form describing the open evolution of the machine density operator, $\rho_\mathrm{m}$, subjected to both the effects of the two thermal reservoirs and the qubit's tape (see Appendix A and Ref.~\cite{Manzano:2019}). In the interaction picture, it reads:
\begin{align} \label{eq:master}
 \dot{\rho}_\mathrm{m} &= - i~[V, \rho_\mathrm{m}] + \gamma_\downarrow^\mathrm{q} \mathcal{D}[\sigma_\mathrm{c}^{+} \sigma_\mathrm{h}^{-}]\rho_\mathrm{m} + \gamma_\uparrow^{\mathrm{q}} \mathcal{D}[\sigma_\mathrm{c}^{-} \sigma_\mathrm{h}^{+}]\rho_\mathrm{m} \nonumber \\ &~~~~+ \mathcal{L}_\mathrm{c}(\rho_\mathrm{m}) + \mathcal{L}_\mathrm{h}(\rho_\mathrm{m}) \equiv \mathcal{L}_\mathrm{tot} (\rho_\mathrm{m}), 
\end{align}
where the first line describes the effects of the collisions with the tape, and the second line the interaction with cold and hot reservoirs, with:
\begin{equation} \label{eq:lindbladians}
 \mathcal{L}_i (\rho_\mathrm{m}) = \gamma_\downarrow^{i} \mathcal{D}[\sigma_i^{-}]\rho_\mathrm{m} + \gamma_\uparrow^i \mathcal{D}[\sigma_i^{+}]\rho_\mathrm{m},
\end{equation}
the local thermal Lindbladian acting on qubit $i=\mathrm{c},\mathrm{h}$ of the machine, and  $\mathcal{D}[L]\cdot \equiv L \cdot L^\dagger - \frac{1}{2}\{L^\dagger L, \cdot \}$ denoting the usual Lindblad dissipators~\cite{Breuer:2002,Wiseman:2010} with rates fulfilling local detailed balance $\gamma_\downarrow^{i}= \gamma_\uparrow^{i} e^{\beta_i E_i}$.

As can be appreciated in the first line of Eq.~\eqref{eq:master}, the collisions with the tape qubits generate both coherent (driving-like) contributions, as given by the operator $V = r \phi (\sigma_\mathrm{c}^{+} \sigma_\mathrm{h}^{-} c^\ast + \sigma_\mathrm{c}^{-} \sigma_\mathrm{h}^{+} c)$, and dissipative terms analogous to the ones induced by the baths in Eq.~\eqref{eq:lindbladians}, with rates $\gamma_{\uparrow}^\mathrm{q} = r \phi^2 p_1$ and $\gamma_{\downarrow}^\mathrm{q} = r \phi^2 p_0$. These two contributions correspond, respectively, to coherent and incoherent transitions between the inner levels $\ket{10}_\mathrm{m} \leftrightarrow \ket{01}_\mathrm{m}$ of the machine (see Fig.~\ref{fig:2}), and depend explicitly on the initial state of the qubits in the tape, $\rho_\mathrm{q}$. As also noticed in Refs.~\cite{Manzano:2019,Landi:2019}, the coherent contribution appears only when the tape qubits are initialized with non-zero coherence in the energy basis, $|c| \neq 0$ in Eq.~\eqref{eq:rho}. 

On the other hand, the local transitions induced by the baths on the machine qubits, can be written for the case of bosonic reservoirs as $\gamma_\downarrow^i = \Gamma_0^i (N_\mathrm{th}^i + 1)$ and $\gamma_\uparrow^i = \Gamma_0^i N_\mathrm{th}^i$, with $\Gamma_0^i$ the spontaneous emission rate and $N_\mathrm{th}^i = (e^{\beta_i E_i} - 1)^{-1}$ the average number of excitations in the bath at temperature $\beta_i$ with energy $E_i$~\cite{Breuer:2002,Wiseman:2010}. For the ease of simplicity, we assume in the following equal spontaneous emission rates, $\Gamma_0 \equiv \Gamma_0^\mathrm{c}= \Gamma_0^\mathrm{h}$. 

The relative weights of local thermal dissipation and collisional dynamics are then determined by the interplay of the parameters $r$, $\phi$ and $\Gamma_0$. More precisely, for $r \phi^2 \gg \Gamma_0$ the collisional tape qubits becomes the dominant contribution ot the dynamics, spoiling thermal effects induced by the baths. On the other hand, the regime $r \phi^2 \ll \Gamma_0$, corresponds to a weak impact of the tape qubits on the machine, where the incoherent terms in the first line of Eq.~\eqref{eq:master} become negligible (but not necessarily first coherent term proportional to $r \phi$). In the following sections we consider both cases where $r \phi^2 \sim \Gamma_0$ and $r \phi^2 \ll \Gamma_0$ with still $\Gamma_0 < r \phi$.

The validity of the master equation~\eqref{eq:master} is warranted in the case of weak coupling to the local reservoirs, $\Gamma_0 \ll E_i$, as well as weak collisions $\phi \ll 1$ occurring on a fast time-scale with respect to the relaxation time-scales induced by the baths, $\Gamma_0 \ll 1/\tau$. It is indeed this later assumption that allows us to safely split the machine dynamics into pieces coming from the collisions with the tape in one side, and from the local thermal reservoirs in the other side (see App. A).

\section{Steady-state dynamics and thermodynamics} \label{sec:steady}

We focus on the operation of the machine in the stationary regime, namely, after which a sufficient number of collisions with the tape have already taken place. The steady state of the machine in the long time run, $\pi_\mathrm{m} = \lim _{t \to \infty} \rho_m$, can be obtained from Eq.~\eqref{eq:master} by imposing $\mathcal{L}(\pi_\mathrm{m}) = 0$ (see Appendix B) and is of the form:
\begin{equation} \label{eq:ss}
\pi_\mathrm{m} = 
 \begin{pmatrix}
\pi_{00} & 0 & 0 & 0  \\
0 & \pi_{10} & \pi_c & 0  \\
0 & \pi_c^\ast & \pi_{01} & 0 \\
0 & 0 & 0 & \pi_{11}
\end{pmatrix}.
\end{equation}
Crucially, the initial coherence in the tape qubits is partially transferred to the machine in the steady state, that acquires coherence in the non-degenerate subspace $\{ \ket{10}_\mathrm{m}, \ket{01}_\mathrm{m} \}$ of $H_\mathrm{m}$. This is contrast to other models that acquire steady-state coherence only in a degenerate, or nearly-degenerate subspace~\cite{Linden:2010,Scully:2011}.

Under steady-state conditions, the effect of the coherent machine on every incoming qubit of the tape is equivalent and can be described through the following completely positive and trace preserving (CPTP) map 
$\rho_\mathrm{q} \rightarrow \mathcal{E}(\rho_\mathrm{q})$, with:
\begin{align} \label{eq:map}
\mathcal{E}(\rho_\mathrm{q}) =&~ \rho_\mathrm{q} - i \phi [\sigma_\mathrm{q}^{-} \pi_c^\ast + \sigma_\mathrm{q}^{+} \pi_c, \rho_\mathrm{q}] \\ 
&+ \phi^2 \pi_{01} \mathcal{D}[\sigma_\mathrm{q}^{-}]\rho_\mathrm{q} + \phi^2 \pi_{10} \mathcal{D}[\sigma_\mathrm{q}^{+}]\rho_\mathrm{q},  \nonumber 
\end{align}
where again coherent and incoherent contributions can be identified, in analogy to Eq.~\eqref{eq:master}. 
Notice that here, contrary to other models of collisional reservoirs~\cite{Strasberg:2017,Landi:2019}, we distinguish between the initial and final states of the qubits in the tape, which are collected after interaction with the machine (see Fig.~\ref{fig:1}).

The map $\mathcal{E}$ in Eq.~\eqref{eq:map} depends on the machine steady-state density operator elements $\pi_{10}$, $\pi_{01}$ and $\pi_c$, which are linked, at the same time, to the initial state $\rho_\mathrm{q}$ through the coefficients in Eq.~\eqref{eq:master}. Although this intricate dependence, we observe that this map turns out to be contractive, and shows a single fixed point, $\mathcal{E}(\tau_\mathrm{v}) = \tau_\mathrm{v}$, with thermal Gibbs form at the virtual temperature:
\begin{equation} \label{eq:equilibrium}
 \tau_\mathrm{v} = \frac{e^{-\beta_\mathrm{v} {H_\mathrm{q}}}}{Z_\mathrm{q}},
\end{equation}
where $Z_\mathrm{q} = \tr[e^{-\beta_\mathrm{v} {H_\mathrm{q}}}]$, and $\beta_\mathrm{v}$ is the virtual temperature in Eq.~\eqref{eq:virtual}. Whenever the qubits in the tape are prepared in the state~ \eqref{eq:equilibrium} the machine is unable to perform any change on it. Later on we will see that this state indeed corresponds to an equilibrium point.

With the help of Eq.~\eqref{eq:map}, the master equation~\eqref{eq:master}, and the expressions for the steady-state density operator in Eq.~\eqref{eq:ss}, we can directly compute the steady-state energy and heat currents into the qubit's tape and from the baths to obtain:
\begin{align} \label{eq:currents}
 \dot{E}_\mathrm{tape} &= r \tr[H_\mathrm{q}\big(\mathcal{E}(\rho_\mathrm{q}) - \rho_\mathrm{q}\big)] = E_\mathrm{q} (\Delta + \zeta), \nonumber \\
 \dot{Q}_\mathrm{c} &= \tr[H_\mathrm{m} \mathcal{L}_1(\pi)] = - E_\mathrm{c} (\Delta + \zeta), \nonumber \\
 \dot{Q}_\mathrm{h} &= \tr[H_\mathrm{m} \mathcal{L}_2(\pi)] = E_\mathrm{h} (\Delta + \zeta), 
 \end{align}
[$\Delta$ and $\zeta$ are given below in Eq.~\eqref{eq:quantities}] which verify $\dot{E}_\mathrm{tape} = \dot{Q}_\mathrm{c} + \dot{Q}_\mathrm{h}$ owing to Eq.~\eqref{eq:resonance}, that is, any increase of energy in the tape qubits is due to input heat from hot and cold baths. Moreover, the currents verify the relation:
 \begin{equation}
  \frac{\dot{E}_\mathrm{tape}}{E_\mathrm{q}} = - \frac{\dot{Q}_\mathrm{c}}{{E}_\mathrm{c}} = \frac{\dot{Q}_\mathrm{h}}{E_\mathrm{h}},
 \end{equation}
implying that the proportionality of the steady-state currents flowing through the machine only depends on the energy spacings $E_\mathrm{q}$, $E_\mathrm{c}$, and $E_\mathrm{h}$.

In the above expressions~\eqref{eq:currents} we conveniently introduced the two following key quantities
 \begin{align} \label{eq:quantities}
  \Delta \equiv r \phi^2 \big(\pi_{01} p_0 -  \pi_{10} p_1 \big),~~ \zeta \equiv i r \phi \big(\pi_c^\ast c -  \pi_c c^\ast \big),
 \end{align}
which capture the incoherent ($\Delta$) and coherent components ($\zeta$) of the energy currents. The quantity $\zeta \geq 0$ is real and positive, and becomes zero whenever the initial state of the qubits in the tape is diagonal in the energy basis, $c=0$ in Eq.~\eqref{eq:rho}. Instead $\Delta$ can be either positive or negative, and both quantities vanish $\Delta = \zeta = 0$ when $\rho_{\mathrm{q}} = \tau_\mathrm{v}$, implying zero currents in Eq.~\eqref{eq:currents}.

The change in the entropy of the tape due to the interaction with the machine in the steady state can be calculated from Eq.~\eqref{eq:map} using perturbation theory (see Appendix B for details) and is given by:
\begin{align} \label{eq:entropy}
& \dot{S}_\mathrm{tape} = - r \big( \tr[\mathcal{E}(\rho_\mathrm{q}) \ln \mathcal{E}(\rho_\mathrm{q})] - \tr[\rho_\mathrm{q} \ln \rho_\mathrm{q}] \big)  \\ 
& = (\lambda_+ - \lambda_{-})\ln\left(\frac{\lambda_-}{\lambda_+}\right) \Big( r \phi^2 |\pi_c|^2 \nonumber 
 + \frac{\Delta(p_1 - p_0) - N |c|^2}{(p_1 -p_0)^2 + 4|c|^2} \Big), \nonumber
\end{align}
where $N \equiv r \phi^2 (\pi_{01} + \pi_{10})$ and we denoted the eigenvalues of $\rho_\mathrm{q}$ by $\lambda_{\pm} \equiv (1 \pm \sqrt{(p_1-p_0)^2 + 4|c|^2})/2$. Using the energy and entropy currents, we can now define the nonequilibrium free energy current of the qubits in the tape with respect to temperature $T_\mathrm{c}$,
\begin{equation} \label{eq:free}
\dot{F}_\mathrm{tape} \equiv \dot{E}_\mathrm{tape} - k_B T_\mathrm{c} \dot{S}_\mathrm{tape}. 
\end{equation}
The nonequilibrium free energy characterizes the maximum extractable work from nonequilibrium states by using local unitary operations and contact with a thermal bath (see e.g. Refs.~\cite{Popescu:2014,Parrondo:2015,Gallego:2016}) at the cold temperature $T_\mathrm{c}$~\cite{note1}. Here it plays a prominent role for quantifying the tape qubits as a resource. Importantly, the free energy in Eq.~\eqref{eq:free} takes into account both thermal populations and coherence in the qubits, and can be split into two components by using the relative entropy of coherence~\cite{Janzing:2006,Lostaglio:2015,Landi:2019} (see App. B for details). In Ref.~\cite{Manzano:2019}, it has been shown that this model is able to amplify the relative entropy of coherence of the incoming qubits for an adequate choice of the parameters. Here we instead focus on the ability of this machine to perform thermodynamic tasks that may be enhanced or assisted by coherence.

From Eqs.~\eqref{eq:currents} and \eqref{eq:entropy}, we are now in a position to state the second law of thermodynamics in steady-state conditions as the positivity of the total entropy production rate~\cite{Manzano:2018,Landi:2020}:
\begin{equation} \label{eq:stot}
 \dot{S}_\mathrm{tot} = \dot{S}_\mathrm{tape} - \beta_\mathrm{c} \dot{Q}_\mathrm{c} - \beta_\mathrm{h} \dot{Q}_\mathrm{h} \geq 0,
\end{equation}
where $\beta_i \dot{Q}_i$ are the entropy fluxes associated to the heat currents from/to reservoir $i$ and $\dot{S}_\mathrm{tape}$ is entropy flux associated to exchange of energy between the machine and the qubits in the tape, $\dot{E}_\mathrm{tape}$. Reversibility conditions associated to a zero entropy production rate in Eq.~\eqref{eq:stot} are verified when the tape qubits are initialized in the thermal state at the virtual temperature~\eqref{eq:equilibrium}.

Making use of energy conservation to eliminate one of the two heat currents in~\eqref{eq:stot}, and identifying the nonequilibrium free energy in Eq.~\eqref{eq:free}, we can rewrite Eq.~\eqref{eq:stot} in the two following equivalent forms:
\begin{align} \label{eq:efficiency}
 \eta &\equiv \frac{\dot{F}_\mathrm{tape}}{\dot{Q}_\mathrm{h}} \leq 1 - \frac{\beta_\mathrm{h}}{\beta_\mathrm{c}}, \\ \label{eq:cop}
 \epsilon &\equiv \frac{\dot{Q}_\mathrm{c}}{-\dot{F}_\mathrm{tape}} \leq \frac{\beta_\mathrm{c}}{\beta_\mathrm{c} - \beta_\mathrm{h}},
\end{align}
where $\eta$ stands for the efficiency of a heat engine increasing the free energy of the tape qubits $\dot{F}_\mathrm{tape}$ using the heat current from the hot bath $\dot{Q}_\mathrm{h}$ as a resource, and $\epsilon$ is the so-called coefficient of performance (COP) of a refrigerator extracting heat from the cold reservoir $\dot{Q}_\mathrm{c}$ by consuming free-energy of the input qubits on the tape. Remarkably, the second-law inequality~\eqref{eq:stot} bounds both quantities, respectively, by Carnot efficiency $\eta_C \equiv 1- \beta_\mathrm{h}/\beta_\mathrm{c}$~\cite{Carnot} and reversible COP, $\epsilon_C \equiv \beta_\mathrm{c}/(\beta_\mathrm{c} - \beta_\mathrm{h})$~\cite{Kosloff:2014}. The bounds are reached, similarly to other models of (incoherent) autonomous thermal machines, at the equilibrium point ($\rho_\mathrm{q} = \tau_\mathrm{v}$), where every energy and entropy currents become zero. Similar notions of nonequilibrium free-energy efficiencies have been also proposed in the context of squeezed thermal reservoirs~\cite{ManzanoPhD}, molecular machines~\cite{Brown:2019}, dissipative chemistry~\cite{Penocchio:2019} and multiterminal mesoscopic conductors~\cite{Hajiloo:2020}.

\section{Regimes of operation}  \label{sec:regimes}

Equations \eqref{eq:efficiency} and \eqref{eq:cop} suggest that the thermal machine introduced here may act either as a heat engine or as a power-driven refrigerator, where the the free energy current to (from) the qubit tape plays the role of the output (input) work. This idea can be made more precise by exploring the joint behavior of the heat currents from the thermal reservoirs $\dot{Q}_\mathrm{c}$ and $\dot{Q}_\mathrm{h}$ and the free energy changes of the qubits in the tape, $\dot{F}_\mathrm{tape}$. We use different parameters of the thermal machine and initial states of the tape qubits, $\rho_\mathrm{q}$. The following three regimes of operation are found:
\begin{enumerate}
\renewcommand{\labelenumi}{\Roman{enumi}}
 \item {\bf Free-energy heat engine (HE):} The thermal machine performs a continuous increase in the free energy of the qubits of the tape, $\dot{F}_\mathrm{tape} \geq 0$, using the natural heat current from the hot bath, $\dot{Q}_\mathrm{h} \geq 0$, which is partially dumped into the cold bath, $\dot{Q}_\mathrm{c} \leq 0$.

 \item {\bf Refrigerator (R):} A heat current from the cold bath is extracted $\dot{Q}_\mathrm{c} \geq 0$ by consuming free energy of the tape qubits, $\dot{F}_\mathrm{tape} < 0$, while heat in transferred into the hot bath $\dot{Q}_\mathrm{h} < 0$. This configuration can be also seen as a free-energy driven heat pump.

 \item {\bf Dissipator (D):} An input free-energy current from the tape qubits $\dot{F}_\mathrm{tape} <0$ is, together with a heat current from the hot bath, $\dot{Q}_\mathrm{h} > 0$ dumped into the cold bath, $\dot{Q}_\mathrm{c} < 0$. In this regime no particularly useful thermodynamic tasks is performed.
\end{enumerate}

The simplicity of the model allows us to obtain analytically the boundaries among steady-state regimes of operation in the absence of initial coherence in the tape. Taking $c=0$ in Eq.~\eqref{eq:rho} immediately yields $\zeta=0$ in Eqs.~\eqref{eq:currents}. Moreover we obtain a purely classical free energy, $\dot{F}_\mathrm{tot} = E_\mathrm{q} \Delta (1- \beta_\mathrm{q}/\beta_\mathrm{c})$, where $\beta_\mathrm{q} \equiv \log(p_0/p_1)/E_\mathrm{q}$ represents the temperature of the incoming qubits in the tape. As a consequence, the regimes of operation in this case are solely determined by the interplay between $\beta_\mathrm{q}$, the virtual temperature $\beta_\mathrm{v}$, and the temperature of the cold bath $\beta_\mathrm{c}$. We recall that here $\beta_\mathrm{v} \leq \beta_\mathrm{c}$ by construction. 
The refrigerator regime R is obtained when the qubits in the tape have initial temperatures higher than the virtual one (absorption refrigerator), or equivalently, when they show a higher population inversion (power-driven refrigerator), $\beta_\mathrm{q} \leq \beta_\mathrm{v}\leq \beta_\mathrm{c}$. This implies $\Delta \leq 0$ in Eq.~\eqref{eq:currents}, leading to $\dot{Q}_{\mathrm{c}} \geq 0$, $\dot{Q}_\mathrm{h} \leq 0$ and $\dot{E}_\mathrm{tape} \leq 0$. On the other hand, when the largest temperature (or largest negative temperature) is the virtual temperature, $\beta_\mathrm{v} \leq \beta_\mathrm{q} \leq \beta_\mathrm{c}$, the energetic currents change sign, $\Delta \geq 0$, and the machine is able to pump free energy into the tape, $\dot{F}_\mathrm{tape} \geq 0$, hence obtaining the heat engine regime HE. Finally, for temperatures of the tape qubits lower than the cold bath, $\beta_\mathrm{v} \leq \beta_\mathrm{c} \leq \beta_\mathrm{q}$, while we still have energy pumping into the tape (since $\Delta > 0$), the entropy flux associated to that energy current, $\dot{S}_\mathrm{tape}$ in Eq.~\eqref{eq:entropy} becomes so high that produces a loss of free energy in the tape, $\dot{F}_\mathrm{tape} \leq 0$, spoiling work extraction and leading to the (useless) dissipator regime D.

\begin{figure}[t!]
\includegraphics[width=1.0 \linewidth]{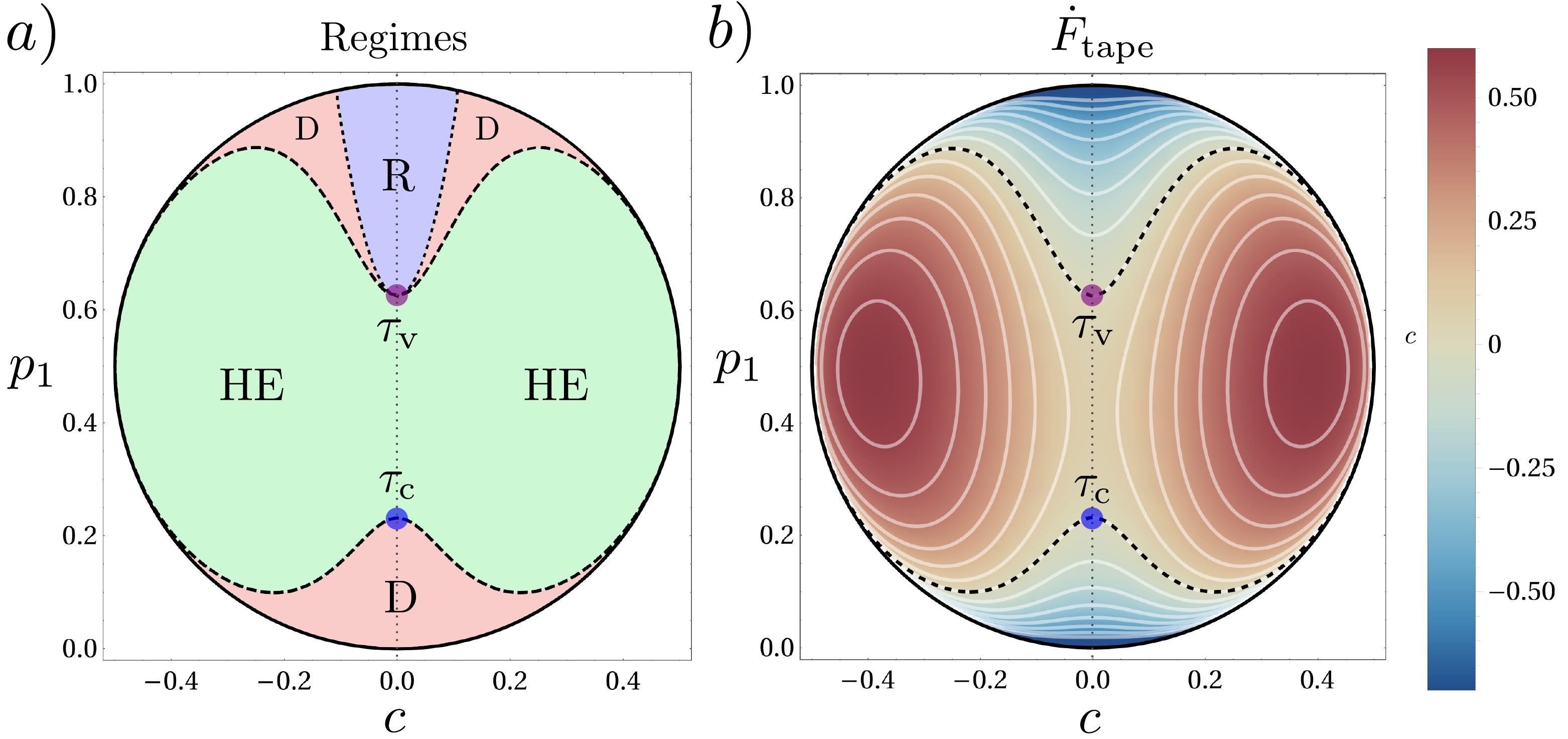}
\caption{(a) Achievable regimes of operation as a function on the initial tape qubits excited population $p_1$ and off-diagonal element $c$, for the case of a relatively large temperature gradient, $\beta_\mathrm{h} = 0.05 \beta_\mathrm{c}$. (b) Corresponding free energy changes in the tape qubits, in units of $r \phi^2 E_\mathrm{q}$. The dashed line represent $\dot{F}_\mathrm{tape} = 0$ and the purple and blue dots in both plots indicate the thermal states $\tau_\mathrm{v}$ and $\tau_\mathrm{c}$, respectively. Parameters: $E_\mathrm{c} = 0.5 E_\mathrm{q}$, $E_\mathrm{h} = 1.5 E_\mathrm{q}$, $\beta_\mathrm{c} = 1.2/E_\mathrm{q}$, $r = 2/E_\mathrm{q}$, $\phi= 0.02$ and $\Gamma_0 = 0.0025/E_\mathrm{q}$.}
\label{fig:regimesI}
\end{figure}

\begin{figure}[t!]
\includegraphics[width=1.0 \linewidth]{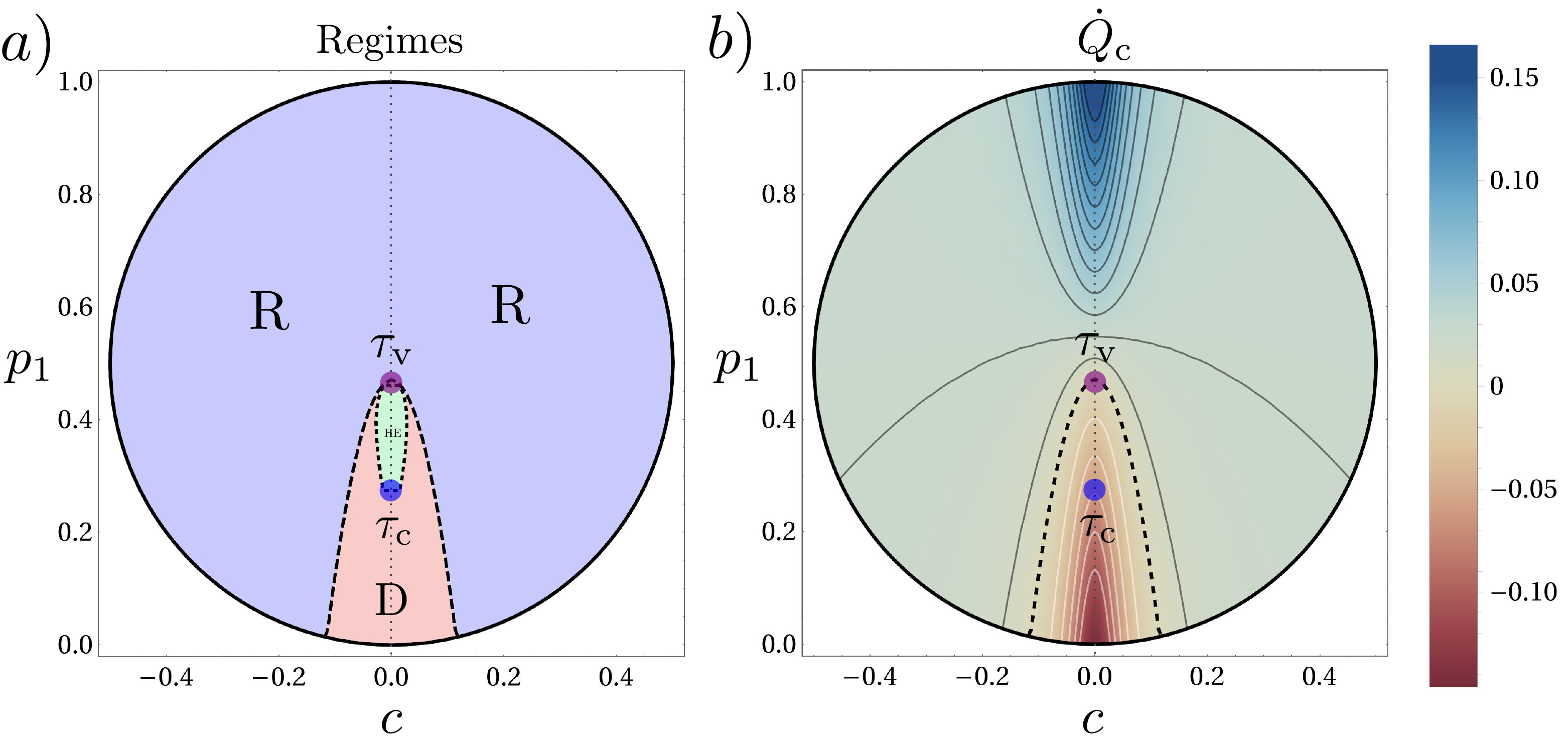}
\caption{(a) Regimes of operation as a function of the initial tape qubits excited population $p_1$ and off-diagonal element $c$, for the case of a relatively small temperature gradient, $\beta_\mathrm{h} = 0.5 \beta_\mathrm{c}$. (b) Cooling power associated to the regimes in (a) in units of $\Gamma_0 E_\mathrm{c}$. The dashed line represent $\dot{Q}_\mathrm{c} = 0$. Again purple and blue dots mark thermal states $\tau_\mathrm{v}$ and $\tau_\mathrm{c}$, respectively. Parameters: $E_\mathrm{c} = 0.8 E_\mathrm{q}$, $E_\mathrm{h} = 1.5 E_\mathrm{q}$, $\beta_\mathrm{c} = 1.2/E_\mathrm{q}$, $r = 2/E_\mathrm{q}$, $\phi= 0.04$ and $\Gamma_0 = 0.0025/E_\mathrm{q}$.}
\label{fig:regimesII}
\end{figure}

A more complex situation arises when the input atoms have a non-zero initial coherence, $c \neq 0$ in Eq.~\eqref{eq:rho}. In that case an explicit analytical solution for the regimes boundaries is not available and we need to rely on numerical evaluation of the currents. In Figs.~\ref{fig:regimesI} and \ref{fig:regimesII} we show, for different representative sets of parameters, the regimes of operation achieved by the machine as a function of the tape qubits initial state, $\rho_\mathrm{q}$. There we represent a cross section of the incoming qubit's Bloch sphere (in interaction picture) and, since the results only depend on $|c|$, we took $c$ to be a real number without loss of generality.

The inclusion of coherence in the tape can greatly modify the regimes of operation with respect to the incoherent case (vertical dotted line at $c=0$). The parameters in Fig.~\ref{fig:regimesI}(a) corresponds to a relatively big temperature gradient in the baths, $\beta_\mathrm{h} = 0.05 \beta_\mathrm{c}$, which favours a broad heat engine regime (green area) with respect to refrigeration (blue area) and the dissipator (red area). We also take a small value of the effective coupling of the tape qubits $\phi = 0.02$ ($r\phi^2 < \Gamma_0$). In Fig.~\ref{fig:regimesI}(b) we show the corresponding free energy changes in the tape qubits, $\dot{F}_\mathrm{tape}$, in units of $E_\mathrm{q} r \phi^2$. Remarkably, we observe that the output free energy current can be enhanced up to 6 times with respect to its maximum incoherent value ($\dot{F}_\mathrm{tape} \simeq 0.1 E_\mathrm{q} r \phi^2$) for values of the initial coherence in the tape qubits $|c| \simeq 0.4$ and $p_1 \simeq 0.5$. 

On the other hand, Fig.~\ref{fig:regimesII}(a) corresponds to a smaller temperature gradient between the baths, $\beta_\mathrm{h} =0.5 \beta_\mathrm{c}$ leading to a wider R regime (blue area), at expenses of the dissipator D (red area) and heat engine HE regimes (tiny green area). We also consider a higher effective coupling, $\phi = 0.04$, leading to a stronger impact of the tape qubits on the machine dynamics than the thermal reservoirs. Since the fuel used for cooling is provided by the free energy from the tape, a higher impact of the later ($r \phi^2 > \Gamma_0$) helps to improve the cooling power.
The heat current from the cold reservoir $\dot{Q}_\mathrm{c}$ (or cooling power) is plotted in Fig.~\ref{fig:regimesII}(b) in units of $E_\mathrm{c} \Gamma_0$. In this case we observe no enhancement in the cooling power by using initial coherence in the tape qubits. However, enhancements can be found for other sets of parameters (see e.g. Fig.~\ref{fig:optimizationII}). In any case, we observe a large region of parameters (south hemisphere of the Bloch sphere) where refrigeration becomes possible owing to the input coherence. In this region we therefore find a coherence-powered refrigerator. In all plots the blue and purple dots represent respectively the thermal states of the qubits with respect to $\beta_\mathrm{c}$, that is, $\tau_\mathrm{c} \equiv e^{-\beta_\mathrm{c} H_\mathrm{q}}/Z_\mathrm{c}$, and $\beta_\mathrm{v}$, that is, $\tau_\mathrm{v}$ in Eq.~\eqref{eq:equilibrium}.

We notice that since the coherent component of the currents in Eq.~\eqref{eq:currents} is non-negative, $\zeta \geq 0$, an enhancement in the heat and energy currents with respect to the incoherent case may be expected. However, we remark that Figs.~\ref{fig:regimesI} and \ref{fig:regimesII} show not only the possibility of coherence-induced enhancements in some regimes, but also that, due to the non-monotonic behavior of the currents with $c$, some thermodynamic operations become enabled by means of coherence in otherwise impossible situations and, for other cases, coherence maybe detrimental. 

\begin{figure}[tb]
\includegraphics[width=0.92 \linewidth]{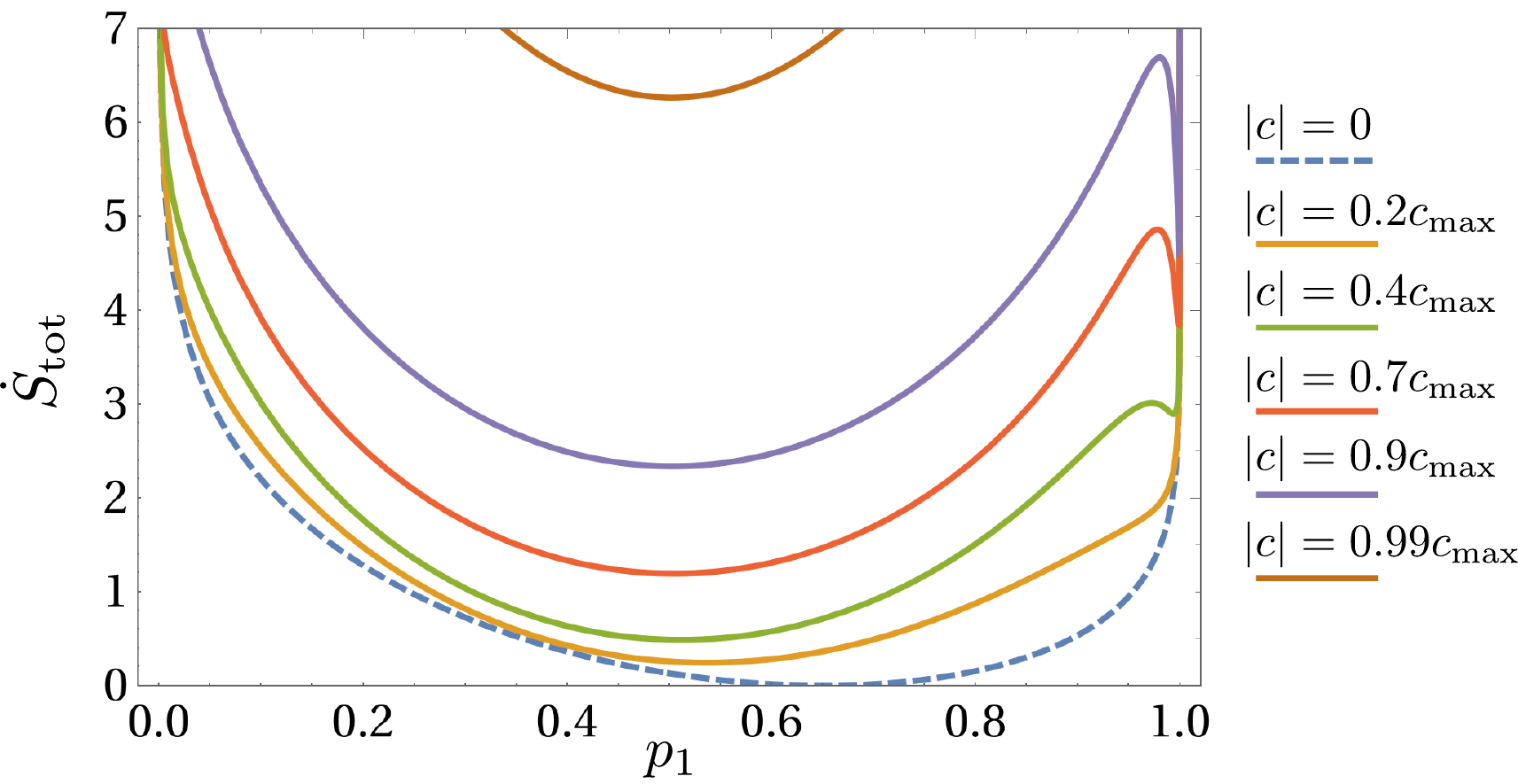}
\caption{Entropy production rate as a function of the tape qubits population $p_1$ for different values of the off-diagonal element absolute vale $|c|$ (see legend). The dashed line corresponds to the incoherent case. Parameters: $E_\mathrm{c} = 0.6 E_\mathrm{q}$, $E_\mathrm{h} = 1.6 E_\mathrm{q}$, $\beta_\mathrm{c} = 1.2/E_\mathrm{q}$, $\beta_\mathrm{h} = 0.05 \beta_\mathrm{c}$, $r = 2.5/E_\mathrm{q}$, $\phi= 0.08$ and $\Gamma_0 = 0.0025/E_\mathrm{q}$.}
\label{fig:ep}
\end{figure}

An important observation at this point is that, by employing initial coherence in the tape qubits, the distance of $\rho_\mathrm{q}$ from the equilibrium state $\tau_\mathrm{v}$ increases in general, forcing the machine to work in further from equilibrium conditions. In other words, increasing the heat and energy currents comes at the price of increasing dissipation, which may eventually spoil the efficiency in the enhanced regimes. It is hence of primary interest to take into account the level of irreversibility of the machine operation, and, ultimately, compare the enhancements in power with respect to the efficiency achieved in each regime. While we will explore in detail the trade-off between power and efficiency in the next section, here we concentrate on the impact of coherence in dissipation as measured by the entropy production rate in Eq.~\eqref{eq:stot}. In Fig.~\ref{fig:ep} we show, for a particular choice of parameters, $\dot{S}_\mathrm{tot}$ as a function of $p_1$ for different values of $|c|$. We find a moderate  increase in dissipation when increasing $|c|$ in the region close to the equilibrium point, at $p_1 \simeq 0.65$. However, as $|c|$ gets close to its maximum value $c_\mathrm{max} \equiv \sqrt{p_0 p_1}$, the entropy production starts growing exponentially. This is more apparent when we get closer to the maximum and minimum values of $p_1$, corresponding to small values of $c_\mathrm{max}$.

Finally, it is also worth pointing out that here we used the free energy in Eq.~\eqref{eq:free} to provide a notion of useful work performed by the thermal machine, since it naturally follows from the formulation of the second law in Eqs.~\eqref{eq:stot} and \eqref{eq:efficiency}. However, this interpretation is not the only one possible. For example one may be interested in alternative notions of work, applicable when assuming extra restrictions in the setup, see e.g. Refs.~\cite{Lostaglio:2015,Niedenzu:2019}. One interesting alternative is given by the so-called \emph{ergotropy}, defined as the maximum extractable work from a quantum state by only using unitary operations performing a cyclical variation of Hamiltonian parameters~\cite{Allahverdyan:2004}. The ergotropy is strictly upper-bounded by the free energy for any temperature~\cite{Zambrini:2018}, and, like the nonequilibrium free energy, can be split into coherent and thermal components~\cite{Francica:2020}. In appendix C we show that similar results as those reported here are obtained when replacing the free energy changes in the tape qubits [Eq.~\eqref{eq:free}] by the change in their ergotropy, as induced by the machine operation. 

\begin{figure*}[t]
\includegraphics[width=0.95 \linewidth]{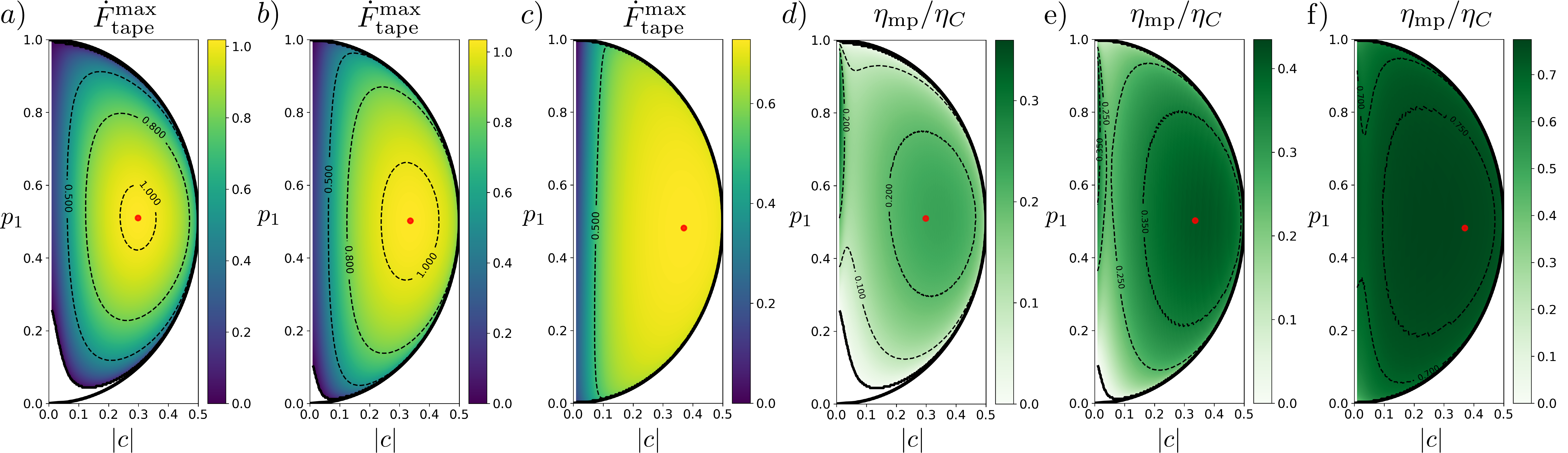}
\caption{(a)-(c) Maximum free energy production $\dot{F}_\mathrm{tape}^\mathrm{max}$ in $r\phi^2 E_\mathrm{q}$ units for temperature scales $\beta_\mathrm{c} E_\mathrm{q} = \{1, 2, 10\}$ respectively and (d)-(f) efficiency at maximum power divided by Carnot efficiency, $\eta_\mathrm{mp}/\eta_C$, for the same temperatures, as a function of the tape qubits initial state parameters $p_1$ and $|c|$. The red dots indicate the absolute maximum of $\dot{F}_\mathrm{tape}^\mathrm{max}$. In the white region inside the semi-circumference the HE regime is not achievable within the range of values used for the optimization, $E_\mathrm{m} \in [E_\mathrm{q},20 E_\mathrm{q}]$. Parameters: $\beta_\mathrm{h} = 0.05 \beta_\mathrm{c}$, $r = 2/E_\mathrm{q}$, $\phi= 0.02$ and $\Gamma_0 = 0.0025/E_\mathrm{q}$.}
\label{fig:optimizationI}
\end{figure*}

\section{Optimizing the performance} \label{sec:optimization}

The results presented in the previous section raise the question of up to what extent coherence can be employed to improve the two more useful regimes in our configuration, namely, free energy generation in the qubits of the tape or cooling the cold reservoir. More precisely, given fixed thermal resources as determined by the temperatures of the baths $\beta_\mathrm{c}$ and $\beta_\mathrm{h}$, and fixing also the coupling-dependent parameters $\Gamma_0$, $r$ and $\phi$, we ask ourselves: What is the optimal design of the machine ($E_\mathrm{c}$ and $E_\mathrm{h}$) and the optimal initial states of the qubits in the tape ($\rho_\mathrm{q}$)  maximizing the free energy production (cooling power)? Is it possible to reach such maximum power regime at a reasonable efficiency (coefficient of performance)? May energetic coherence lead to an absolute enhancement for fixed resources?

In order to address these questions we set the temperature scale of the setup by enforcing different values of the cold bath temperature $k_B T_\mathrm{c} = \{ 1.0~E_\mathrm{q}, 0.5~E_\mathrm{q}, 0.1~E_\mathrm{q} \}$ and use as a free parameter in the optimization the largest energy spacing in the machine Hilbert space, $E_\mathrm{m} \equiv E_\mathrm{c} + E_\mathrm{h}$. We also fix the bath relaxation rates as $\Gamma_0 = 0.0025/E_\mathrm{q}$, 
and the parameters $r = 2/E_\mathrm{q}$ and $\phi= 0.02$, such that the action of the tape qubits in the machine is weak compared to the thermal reservoirs, $r \phi^2 \ll \Gamma_0$. This implies a weak back-reaction of the tape qubits on the machine, favoring the coherent exchange of energy between machine and tape. Hence we expect a higher impact of coherence in this regime of parameters.

\subsection{Maximization of the free energy}

We numerically maximize the free energy current in the tape qubits by varying the the largest energy gap in the machine, $E_\mathrm{m} \equiv E_\mathrm{c} + E_\mathrm{h}$ in the interval $[E_\mathrm{q}, 20 E_\mathrm{q}]$, for fixed $E_\mathrm{q}$. We denote the optimal value of the free energy current by $\dot{F}_\mathrm{tape}^\mathrm{max}$. Since here we are interested in the performance of the machine as a heat engine, we consider a relatively big temperature gradient $\beta_\mathrm{c} = 0.05 \beta_\mathrm{h}$, leading to a value for Carnot efficiency $\eta_C = 0.95$. 

We show the results for the optimized free energy production $\dot{F}_\mathrm{tot}^\mathrm{max}$ as a function of the initial state of the tape qubits in Fig.~\ref{fig:optimizationI}(a)-(c). Since all results are symmetric with respect to the axis $c=0$, we only show half of the Bloch sphere section. The three plots correspond to the three choices of $\beta_\mathrm{c}$ as stated above, i.e. moderate, low, and extremely-low temperatures, respectively. In all three cases, we obtain a clear maximum at values of $p_1$ around $0.5$ and initial coherence $|c|$ between $0.3$ and $0.37$ (see red dots). This is an important feature, which indicates that the heat engine achieves maximum power when it is assisted by energetic coherence in the tape qubits. The free energy generation $\dot{F}_\mathrm{tape}^\mathrm{max}$ is enhanced with respect to its maximum value without coherence ($|c|=0$) about $4.76$ times in Fig.~\ref{fig:optimizationI}(a), $6.1$ times in Fig.~\ref{fig:optimizationI}(b) and $3.34$ times in Fig.~\ref{fig:optimizationI}(c). The corresponding values of $E_\mathrm{m}$ optimizing free energy generation are, respectively, $7.3 E_\mathrm{q}$, $3.9 E_\mathrm{q}$, and $1.7 E_\mathrm{q}$, and vary only very slightly wihin the yellowish areas in Fig.~\ref{fig:optimizationI}.
We also notice that optimization over $E_\mathrm{m}$ allows broadening the set of initial states $\rho_\mathrm{q}$ for which the HE regime is obtained. More precisely, the regions close to the north and south poles of the Bloch sphere for which operations R and D are achieved [see Fig.~\ref{fig:regimesI}] eventually become HE by allowing larger values of $E_\mathrm{m}$. The white areas close to the south pole in Figs.~\ref{fig:optimizationI}(a) and (b) correspond to parameters for which the HE regime can only be achieved by increasing $E_\mathrm{m}$ over the maximum value $20 E_\mathrm{q}$ considered in the optimization procedure. On the other hand, for very low temperatures [Fig. \ref{fig:optimizationI}(c)] the HE regime is achieved for values $E_\mathrm{m} < 2.7 E_\mathrm{q}$ within the whole Bloch sphere.

\begin{figure*}[t]
\includegraphics[width=0.95 \linewidth]{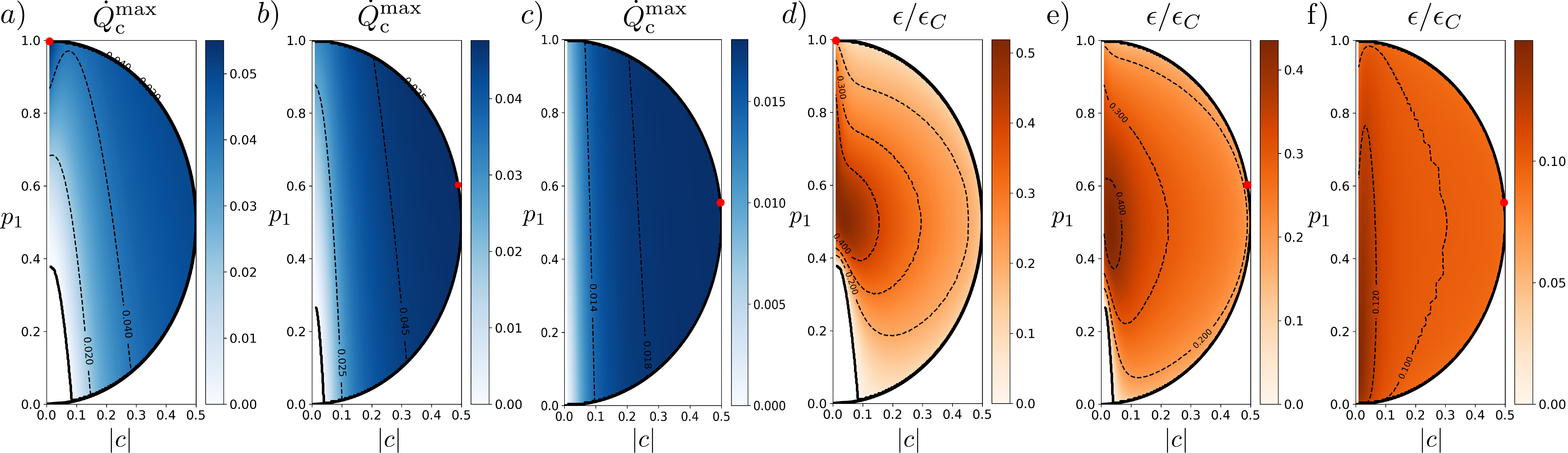}
\caption{(a)-(c) Maximum cooling power $\dot{Q}_\mathrm{c}^\mathrm{max}$ in $\Gamma_0 E_\mathrm{q}$ units  and (d)-(f) COP at maximum power divided by the reversible COP, $\eta/\eta_C$, for temperature scales $\beta_\mathrm{c} E_\mathrm{q} = \{1, 2, 10\}$, as a function of the tape qubits initial state parameters $p_1$ and $|c|$. The red dots indicate the absolute maximum of $\dot{Q}_\mathrm{c}^\mathrm{max}$. White areas inside the semi-circumference denote parameters regions for which the refrigerator regime R cannot be achieved. Parameters: $\beta_\mathrm{h} = 0.5 \beta_\mathrm{c}$, $r = 2.5/E_\mathrm{q}$, $\phi= 0.08$ and $\Gamma_0 = 0.0025/E_\mathrm{q}$.}
\label{fig:optimizationII}
\end{figure*}

In Figs.~\ref{fig:optimizationI}(d)-(f) we show the efficiency at maximum free energy power divided by Carnot's efficiency $\eta_C$ for the three values of the cold bath temperature, $\beta_\mathrm{c}$. For moderate temperatures [Fig.~\ref{fig:optimizationI}(d)] we observe a general detrimental role of coherence. The efficiency peaks at $p_1=0.9$ and $|c|=0$, for which $\eta \simeq 0.47 \eta_C$ and quickly drops to values around $\eta = 0.2 \eta_C$ in the region where maximum free energy power is obtained [see Fig.~\ref{fig:optimizationI}(a)]. The situation radically changes for smaller temperatures of the cold bath [Figs.\ref{fig:optimizationI}~(e) and (f)], where a second peak in the efficiency emerges at $p_1 \sim 0.5$ and $|c|\sim 0.3$ leading to efficiencies around $0.35 \eta_C - 0.42 \eta_C$ in the region of maximum free energy generation in Figs.~\ref{fig:optimizationI}(b) and (c). These efficiencies are comparable to the ones reached for the first peak ($\eta \simeq 0.44 \eta_C$ at $p_1 \simeq 0.88$ and $|c|=0$).
This second peak broadens when further decreasing the temperature (increasing $\beta_\mathrm{c}$) [see Fig.~\ref{fig:optimizationI}(f)] reaching high values for the efficiency around $\eta \simeq 0.78 \eta_C$ at maximum free energy output, to be compared to $\eta \simeq 0.75 \eta_C$, the maximum available efficiency at maximum power without using coherence in the tape (reached for $p_1 \simeq 0.73$). This is a remarkable feature, showing that the use and control of coherence at low temperatures plays a crucial role in developing power, and can be used to enhance the performance (both power and efficiency at maximum power) of thermal machines.

\subsection{Maximization of cooling power}

As a second case of interest, we consider the maximization of the cooling power, $\dot{Q}_\mathrm{c}$, that is, the heat current extracted from the cold reservoir by the machine. As before, optimization is performed by varying the largest energy gap in the machine, $E_\mathrm{m} \in [E_\mathrm{q}, 20 E_\mathrm{q}]$ for fixed $E_\mathrm{q}$, and the three temperatures scales $\beta_\mathrm{c} E_\mathrm{q} = \{1, 2, 10 \}$. On the contrary, in this case we consider a relatively small temperature gradient $\beta_\mathrm{c} = 0.5 \beta_\mathrm{h}$, favoring cooling regimes R with respect to HE ones [see e.g. Fig.~\ref{fig:regimesII}]. 
The reversible (maximum) COP value associated to this temperature ratio is $\epsilon_C = 1$ in all three cases. 

In Figs.~\ref{fig:optimizationII}(a)-(c) we show the maximum cooling power $\dot{Q}_\mathrm{c}^\mathrm{max}$ in units of $\Gamma_0 E_\mathrm{q}$ as a function of the initial state $\rho_\mathrm{q}$ for the three different values of $\beta_\mathrm{c}$. White areas in the bottom represent regions of parameters where cooling cannot be achieved, no matter the value of $E_\mathrm{m}$. The optimal values of $E_\mathrm{m}$ lie in between $E_\mathrm{q}$ and $4.5 E_\mathrm{q}$ in any case, being smaller for lower temperatures of the cold bath. Similarly to the previous case, our results indicate that cooling can be optimized by allowing the initial states of the tape qubits to have non-zero coherence. This can be indeed appreciated in Figs.~\ref{fig:optimizationII}(b) and (c), where the maximum cooling power is achieved in the right part of the semi-cross sections of the tape qubits Bloch sphere. In particular the maximum of $\dot{Q}_\mathrm{c}^\mathrm{max}$ is achieved at points $(p_1 \simeq 0.6, |c| \simeq 0.49)$ for $\beta_\mathrm{c} = 2$ and at $(p_1 \simeq 0.55, |c| \simeq 0.50)$ for $\beta_\mathrm{c} = 10$, enhancing optimal incoherent cooling by $1.5$ and $2.4$ times, respectively. On the other hand, we confirm that allowing a non-zero coherence in the tape, allows reaching refrigeration in situations where it would be otherwise impossible, that is, for $p_1 \leq 0.38$ in Fig.~\ref{fig:optimizationII}(a), and for $p_1 \leq 0.26$ in Fig.~\ref{fig:optimizationII}(b).  

Figs.~\ref{fig:optimizationII}(d)-(f) show the COP of the fridge $\epsilon$ at maximum cooling power, divided by the reversible (Carnot) value $\epsilon_C$, corresponding to Figs.~\ref{fig:optimizationII}(a)-(c). In this case maximum values of the COP are obtained for $|c|=0$ in all cases of moderate, low and very low temperatures. The overall maxima are located respectively at $p_1 \simeq 0.54$, $p_1\simeq 0.51$ and $p_1 \simeq 0.27$, generating a high-COP area around them. As temperature decreases the values the COP become lower and more homogeneous over the whole Bloch sphere semi-cross section. Here we observe that the areas corresponding to maximum values of the cooling power $\dot{Q}_\mathrm{c}^\mathrm{max}$ and maximum COP do not overlap between them. This means that, contrary to the free energy generation case, coherence-enhanced cooling is not accompanied by an improved efficiency. Therefore, in this case the customary trade-off between power and efficiency extends to the initial state $\rho_\mathrm{q}$ optimization.

\section{Conclusions} \label{sec:conclusions}

We have shown that, using energetic coherence, thermodynamic tasks can be enabled and/or enhanced by autonomous thermal machines. In particular we considered a setup where thermal and coherent resources mix up in a beneficial way. We explored both situations where the tape of qubits is operated by the machine to obtain a useful output (free energy), as well as cases in which the qubits in the tape are used as a fuel to power standard thermodynamic tasks, like refrigeration of a cold reservoir. The relative simplicity of the model allowed us to quantitatively assess the impact of the initial state of the qubits in the tape, $\rho_\mathrm{q}$, and in particular its initial coherence $c$, in these two main situations.

Our results indicate that coherence can notably enhance the power of the machine in some regimes of operation, and that coherence can be used to enable machine operation in otherwise impossible situations. 
These enhancements are not obtained for every value of the initial coherence $c$ in general, but it needs to be properly selected. Allowing optimization over the machine design results in broader regions of parameters where coherence plays a beneficial role. We have shown that these improvements are the result of a nonequilibrium effect, with a non-zero associated entropy production. Our results go beyond weakly-coherent models~\cite{Landi:2019}, linear-response regime~\cite{Brandner:2017} or weak-dissipation engines~\cite{Abiuso:2020}, paving the way for understanding energetic-coherence-enhanced performance in autonomous thermal machines.

The optimization procedure employed here shows that energetic coherence is essential in order to maximize the machine output free energy and cooling power. Furthermore, we observed that in the first case, coherence also enhances the efficiency at optimal operation for sufficiently low temperatures. These promising results may be further explored in experimental one-atom maser configurations~\cite{Muller:1985,Klein:1987,Dotsenko:2020} and extended to other autonomous configurations beyond the case of qubits considered here. For example, it would be interesting to consider situations where the tape of qubits is replaced by other physical systems, like e.g. harmonic oscillators, also in view of possible experimental implementations in trapped ion setups~\cite{Poschinger:2019} or circuit QED architectures~\cite{Pekola:2019}.

\begin{acknowledgments}
 K.H. thanks Prof. Abderrahim El Allati for interesting discussions and providing useful comments. G.M. acknowledges Paul Skrzypczyk, Ralph Silva and Juan M. R. Parrondo for related discussions. This work has been partially supported by ICTP through the ICTP-IAEA STEP program. K.H and Y.H would like also to thank the ICTP Office of External Activities (OEA) for the financial support through AF-14.
 G.M. acknowledges funding from the European Union's Horizon 2020 research and innovation program under the Marie Sk\l{}odowska-Curie grant agreement No 801110 and the Austrian Federal Ministry of Education, Science and Research (BMBWF).
\end{acknowledgments}

\appendix
\section{Derivation of the dynamics}

We obtain the master equation \eqref{eq:master} in Sec.~\ref{sec:model} for the thermal machine by explicitly modelling its interaction with the thermal baths and with the qubits in the tape. In particular we consider bosonic baths with Hamiltonian $H_B^{(i)}=\sum_k \Omega_k^{(i)} a_k^{(i) \dag} a_k^{(i)}$ for $i=\mathrm{c},\mathrm{h}$ and where the index $k$ labels the bath's field modes with frequencies $\Omega^{(i)}_k$, and $a_k^{(i) \dag}(a_k)$ being their usual creation (annihilation) operators. We assume a weak coupling between the baths and the corresponding machine qubits, in the rotating wave approximation:
\begin{align}
H_{\mathrm{m}B}= \sum_{i=\mathrm{c},\mathrm{h}} \sum_k g_k^{(i)}(\sigma_{i}a_k^{(i)\dag}+\sigma_{i}^\dag a^{(i)}_k)
\end{align}
with $g_k^{(i)} \ll E_i$ the coupling strength of the i-$th$ qubit to each of the $k$ modes of the bath $i$, with spectral densities $J_i(\Omega)=\sum_k (g_k^{(i)})^2 \delta(\Omega- \Omega_k^{(i)})/\Omega_k^{(i)}$.

In the absence of interactions with the tape qubits, each machine qubit is independently coupled to a local thermal bath. The reduced dynamics of the machine can then be easily obtained using the Born-Markov and secular approximations~\cite{Breuer:2002,Wiseman:2010}: 
\begin{align} \label{eq:L0}
\dot{\rho}_\mathrm{m} (t) &= \sum_{i = \mathrm{c},\mathrm{h}} \gamma_\downarrow^{i} \mathcal{D}[\sigma_i^{-}]\rho_\mathrm{m} + \gamma_\uparrow^i \mathcal{D}[\sigma_i^{+}]\rho_\mathrm{m} \equiv \mathcal{L}_0 (\rho_\mathrm{m}),
\end{align}
where $ \gamma^{i}_\downarrow= \Gamma_0^{i} (1+ N_\mathrm{th}^{i})$, $\gamma^{i}_\uparrow = \Gamma_0^{i}  N_\mathrm{th}^{i}$ are the rates of emission and absorption processes induced by the thermal baths. Here we assumed an Ohmic behavior of the of the baths around the respective qubit frequencies and no modes at the frequency of the complementary qubit, leading to $J_i(E_j) \simeq (\Gamma_0^{i}/2\pi) \delta_{i j}$.

Then, we consider the successive interactions (or collisions) of the machine with a series of qubits in the tape, one at a time. Crucially we assume that the interactions last a very short time $\tau$, which we assume to be smaller than the timescales of the machine's relaxation dynamics, $\tau \ll 1/\Gamma_0^i$ for $i=\mathrm{c},\mathrm{h}$. In this way, the collisions occur almost instantaneously and can be described by a unitary (energy-preserving) operator acting over the machine and the tape qubits, while the presence of the baths during the collisions becomes negligible. For a collision occurring at time $t$ we have $U(t+\tau, t) = e^{-i H_{int} \tau}$. Therefore, after the collision the joint state of machine and qubit tape (in interaction picture) is given by:
\begin{align}\label{lol}
\rho(t+\tau)&= U(t+\tau, t) ~\rho(t)~ U^\dag(t+\tau,t),
\end{align}
where $\rho(t) = \rho_\mathrm{m}(t)\otimes \rho_q$ is the joint state before interaction, corresponding to an (uncorrelated) product state of the machine at time $t$ and a ``fresh'' qubit in the tape. Notice that we assume all tape qubits to have the same phase when they interact with the machine.

The reduced evolution of the machine is given by a completely positive and trace-preserving (CPTP) map, obtained after partial tracing of the qubit degrees of freedom in Eq.~\eqref{lol}. Using the expression of $H_\mathrm{int}$ in Eq.~\eqref{eq:int} and expanding up to second order in $\phi = g \tau \ll 1$, we obtain: 
\begin{align}\label{eq:Theta}
\rho_\mathrm{m}(t+ \tau) &\equiv \Xi(\rho_\mathrm{m}) = \tr_\mathrm{q}[\rho(t+ \tau)]  \\
&= \rho_\mathrm{m}(t) - i \phi [\sigma_\mathrm{c}^{-} \sigma_\mathrm{h}^{+} c^\ast + \sigma_\mathrm{c}^{+}\sigma_\mathrm{h}^{-} c, \rho_\mathrm{m}(t)] \nonumber \\ 
&+ \phi^2 p_1 \mathcal{D}[\sigma_\mathrm{c}^{-} \sigma_\mathrm{h}^{+}]\rho_\mathrm{m}(t) + \phi^2 p_0 \mathcal{D}[\sigma_\mathrm{c}^{+} \sigma_\mathrm{h}^{+}]\rho_\mathrm{m}(t).  \nonumber
\end{align}
Finally we need to combine the dynamics induced by the baths in Eq.~\eqref{eq:L0} with the instantaneous collisions as described by Eq.~\eqref{eq:Theta}. For this purpose, we assume that the collisions occur at random times following Poissonian statistics with rate $r$. 
The state of the machine at time $t$ after $n$ collisions can then be explicitly computed as:
\begin{eqnarray}\label{rest}
\rho_m^{(n)}(t) = \int_{0}^t ds ~w(t-s) e^{\mathcal{L}_0(t-s)}~\Xi(\rho_m^{(n-1)}(s)),
\end{eqnarray}
where $\rho_m^{(n-1)}(s)$ is the state of the system after $n-1$ collisions at time $s$, and $w(\Delta t)= r e^{-r \Delta t}$ is the so-called waiting-time distribution corresponding to Poisson statistics, which captures the probability that a collision does not happen in the interval $\Delta t$. The last term inside the integral describes a collision at time $s$ through the CPTP map $\Xi$, and the second exponential term describes the evolution of the machine up to the next collision, i.e. the interval of time when the tape is stopped and only the thermal baths act on the machine. Taking the time-derivative of Eq.~\eqref{rest}, summing over $n$, and replacing the expressions for the Liouvillian $\mathcal{L}_0$ in Eq.~\eqref{eq:L0} and the CPTP map $\Xi$ in Eq.~\eqref{eq:Theta}, we recover the master equation \eqref{eq:master} in Sec.~\ref{sec:model}.

\section{Steady-state operation}

The collision of the machine with the tape results in driving the machine to a nonequilibrium steady state $\pi_\mathrm{m}$ in the long time run. To obtain it, we separate the master equation \eqref{eq:master} into a set of coupled differential equations for the relevant populations and coherences of $\rho_\mathrm{m}$, which can be expressed in matrix form as $\dot{P} = M \cdot P$, with the vector $P^T \equiv (\varrho_{00}, \varrho_{01}, \varrho_{10}, \varrho_{11}, \varrho_\mathrm{v}, \varrho_\mathrm{v}^\ast)$ containing the elements $\varrho_{ij} = \langle i j| \rho_\mathrm{m} |ij \rangle_\mathrm{m}$ and $\varrho_\mathrm{v} = \langle 10 | \rho_\mathrm{m} |01 \rangle_\mathrm{m}$ of $\rho_\mathrm{m}$, and the dynamical matrix:
\begin{widetext}
\begin{eqnarray}
M= \begin{bmatrix}
- (\gamma^c_\uparrow +\gamma^h_\uparrow ) &  \gamma^c_\downarrow  &  \gamma^h_\downarrow & 0 & 0 & 0\\
\gamma^c_\uparrow  & -\gamma^q_\uparrow - (\gamma^c_\downarrow + \gamma^h_\uparrow)  &  -\gamma^q_\downarrow & \gamma^h_\downarrow &  -i r \phi c^{*} &   i r c \phi \\ 
  \gamma^h_\uparrow & \gamma^q_\uparrow & -\gamma^q_\downarrow  - (\gamma^h_\downarrow + \gamma^c_\uparrow) &\gamma^c_\downarrow & i r \phi c^{*} & - i r c \phi \\
 0 &   \gamma^h_\uparrow &  \gamma^c_\uparrow & - (\gamma^h_\downarrow +\gamma^c_\downarrow ) & 0 & 0\\
 0 & - i r c \phi  &  i r c \phi & 0 & D & 0 \\
 0 &  ir  c^{*} \phi & - ir  c^{*} \phi &0 &0 & D\\
\end{bmatrix},
\end{eqnarray}
\end{widetext}
where $D=-(\gamma^q_\uparrow + \gamma^q_\downarrow) -\frac{1}{2} (\gamma^c_\downarrow + \gamma^h_\downarrow +\gamma^c_\uparrow +\gamma^h_\uparrow)$. The steady-state elements of the density operator in Eq.~\eqref{eq:ss} are calculated by solving the kernel of the dynamical matrix $M$ above, that is, imposing $M \cdot \Pi = 0$, and obtaining $\Pi^T = (\pi_{00}, \pi_{01}, \pi_{10}, \pi_{11}, \pi_\mathrm{v}, \pi_\mathrm{v}^\ast)$. Notice that in the vectors $P$ and $\Pi$, we excluded all other non-diagonal elements of $\rho_\mathrm{m}$, since they are only subjected to exponential decay (decoherence), and are hence zero in the steady state regime. On the other hand, the CPTP map describing the operation of the machine on the tape qubits in the steady state, is obtained, similarly to Eq.~\eqref{eq:Theta}, by tracing out the machine degrees of freedom in Eq.~\eqref{lol}, that is, $\mathcal{E} (\rho_\mathrm{q}) \equiv \tr_\mathrm{m}[\rho(t+\tau)]$. Expanding up to second order in $\phi$ we recover straightforwardly Eq.~\eqref{eq:map}.  

 The changes in energy and entropy of the tape qubits due to the collisions, as given in Eqs.~\eqref{eq:currents} and \eqref{eq:entropy} in Sec.~\ref{sec:steady}, are calculated using the explicit expression of the map $\mathcal{E}(\rho_\mathrm{q})$ in Eq.~\eqref{eq:map}. The only less straightforward point is the computation of the rate of entropy change in Eq.~\eqref{eq:entropy}, for which we expand the eigenvalues and eigenstates of $\mathcal{E}(\rho_\mathrm{q})$, to second order in $\phi$ as 
$\lambda_k^{\prime} = \lambda_k + \phi \lambda_k^{(1)} + \phi^2 \lambda_k^{(2)}$,
where $\lambda_k$ and $\ket{\lambda_k}_\mathrm{q}$ denote the eigenvalues and eigenstates of $\rho_\mathrm{q}$. The first and second-order contributions to the eigenvalues (and eigenstates) are computed by using perturbation theory for the density operator~\cite{Manzano:2019}. The entropy changes in the tape are then approximated, up to second order in $\phi$, by:
\begin{align}\label{eq:entropyB}
 \dot{S}_\mathrm{tape} &= -r \sum_k (\lambda_k^\prime \ln \lambda_k^\prime  - \lambda_k \ln \lambda_k) \nonumber \\
 &\simeq -r \phi^2 \sum_k \lambda_k^{(2)} \ln \lambda_k,
\end{align}
where we used $\lambda_k^{(1)} = 0$ and for the second-order contribution we have: 
\begin{align} \label{eq:second-order}
\lambda_k^{(2)} =~& \pi_{01} \langle \lambda_k | \mathcal{D}[\sigma_\mathrm{q}^{-}]\rho_\mathrm{q} |\lambda_k \rangle  +  \pi_{10} \langle \lambda_k | \mathcal{D}[\sigma_\mathrm{q}^{+}]\rho_\mathrm{q}    |\lambda_k \rangle \nonumber  \\ 
&- \sum_{l \neq k}(\lambda_{l} - \lambda_{k})|\langle \lambda_l|\sigma_\mathrm{q}^{-} \pi_c^\ast + \sigma_\mathrm{q}^{+} \pi_c|\lambda_k \rangle|^2.
\end{align}
By plugging \eqref{eq:second-order} into \eqref{eq:entropyB}, and operating we obtain the expression for the entropy changes in the tape given in Eq.~\eqref{eq:entropy} of Sec.~\ref{sec:steady}.

Finally, we recall that the non-equilibrium free energy changes of the tape qubits $\dot{F}_\mathrm{tape}$ in Eq.~\eqref{eq:free} can be divided into thermal and coherent parts~\cite{Manzano:2019}. This split is performed by introducing the dephased states in the energy basis, $\bar{\rho}_\mathrm{q} = \sum_j \Pi_j \rho_\mathrm{q} \Pi_j$, with $\Pi_j = \{ \vert 0 \rangle \langle 0 \vert_\mathrm{q}, \vert 1 \rangle \langle 1 \vert_\mathrm{q}\}$ the projectors of the spectral decomposition of $H_\mathrm{q}$. 

By considering the dephased states of input and output tape qubits, we can calculate the part of the qubit entropy changes only due to the populations, obtaining $\dot{\bar{S}}_\mathrm{tape} =  (\Delta + \zeta) \ln(p_0/p_1)$. The associated (classical) free energy changes are then defined as $\dot{\bar{F}}_\mathrm{tape} \equiv \dot{E}_\mathrm{tape} -\dot{\bar{S}}_\mathrm{tape}$. The free energy split then reads \cite{Janzing:2006,Santos:2019,Manzano:2019}:
\begin{equation}
 \dot{F}_\mathrm{tape} = \dot{\bar{F}}_\mathrm{tape} + \dot{C}_\mathrm{tape},
\end{equation}
where $\dot{C}_\mathrm{tape} = \dot{\bar{S}} - \dot{S}$ is the change in relative entropy of coherence, $C(\rho) = S(\rho || \bar{\rho}) = S(\bar{\rho}) - S(\rho) \geq 0$, which is non-negative, monotonic and constitutes a proper measure of coherence~\cite{Bartlett:2007,Streltsov:2017}. We notice that whenever the initial state of the qubits in the tape is diagonal in the energy basis ($c=0$), we have $\dot{C}_\mathrm{tape}=0$, and the free energy reduces to its classical contribution.

\section{Ergotropy generation}

The results provided in Sec.~\ref{sec:regimes} for the generation of free energy in the tape qubits, can be complemented by considering ergotropy as an alternative quantifier for the power of the machine. In this way, the operation of the machine would be to increase or decrease the ergotropy of the tape qubits. The name ergotropy was coined in Ref.~\cite{Allahverdyan:2004} to quantify the maximum work extractable from a generic quantum state $\sigma$ by using unitary transformations $\mathcal{U}$ describing the cyclic variation of a parameter controlling the system Hamiltonian $H$:
\begin{align} \label{eq:ergotropy}
 \mathcal{W}(\sigma) = \tr[H \sigma] - \min_{\mathcal{U}} \tr[H~ \mathcal{U} \sigma \mathcal{U}^\dagger] \nonumber \\
  = \sum_{j k} \lambda_{j} E_k  (|\langle \lambda_j| E_k \rangle|^2 - \delta_{j k}),
\end{align}
where $\lambda_1 \geq \lambda_2 \geq ...$ denote the ordered eigenvalues of $\sigma$ and $E_1 \leq E_2 \leq ...$ the eigenvalues of $H$. In the case of qubits, the optimal unitary $\mathcal{U}$ corresponds to a rotation in the Bloch sphere that transforms the initial state $\sigma$ into a thermal Gibbs state $\tau_\beta \equiv e^{-\beta H}/Z$ with $\beta$ determined by the entropy of the initial state $\sigma$.

Applying the definition Eq.~\eqref{eq:ergotropy} to the input and output states of the tape qubits, and calculating their difference, we obtain:
\begin{align} \label{eq:ergodev}
 \dot{\mathcal{W}}_\mathrm{tape} = \dot{E}_\mathrm{tape} - r \phi^2 \lambda_{-}^{(2)} E_\mathrm{q},
\end{align}
where we used that the optimal unitary transforms the eigenstate of $\rho_\mathrm{q}$ corresponding to highest (lowest) eigenvalue, $\lambda_{+} \geq \lambda_{-}$, to the ground (excited) states [similarly for $\mathcal{E}(\rho_\mathrm{q})$], and used $\lambda_{-}^\prime - \lambda_{-} = \phi^2 \lambda_{-}^{(2)}$. Introducing into Eq.~\eqref{eq:ergodev} the expression for $\lambda_{-}^{(2)}$ as given by Eq.~\eqref{eq:second-order}, and operating we obtain
\begin{align} \label{eq:ergotropy-final}
 \dot{\mathcal{W}}_\mathrm{tape} &= \dot{E}_\mathrm{tape} - E_\mathrm{q}(\lambda_+ - \lambda_{-})\cdot \nonumber \\ 
 & \Big( r \phi^2 |\pi_c|^2 + \frac{\Delta(p_1 - p_0) - N |c|^2}{(p_1 -p_0)^2 + 4|c|^2} \Big).
\end{align}

\begin{figure}[tbh]
\includegraphics[width=1.0 \linewidth]{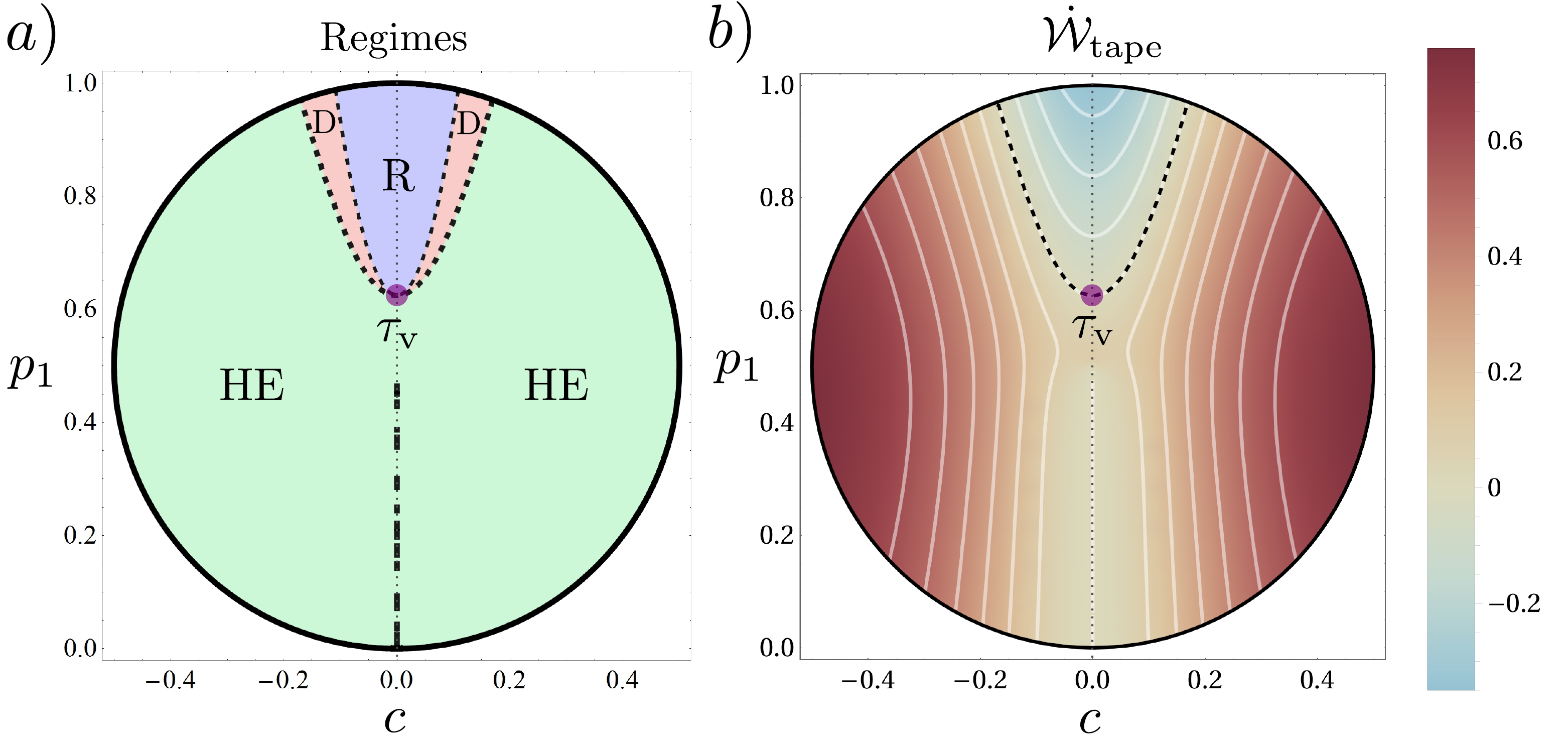}
\caption{(a) Regimes of operation using ergotropy to characterize work in a cross section of the tape qubits Bloch sphere. Ergotropy vanishes along the dashed line at $c=0$ in the south hemisphere. (b) Ergotropy changes in the tape qubits, in units of $r \phi^2 E_\mathrm{q}$. The dashed line represent $\dot{W}_\mathrm{tape} = 0$ and the purple dot is the thermal state $\tau_\mathrm{v}$. Other parameters are as in Fig.~\ref{fig:regimesI}: $\beta_\mathrm{h} = 0.05 \beta_\mathrm{c}$, $E_\mathrm{c} = 0.5 E_\mathrm{q}$, $E_\mathrm{h} = 1.5 E_\mathrm{q}$, $\beta_\mathrm{c} = 1.2/E_\mathrm{q}$, $r = 2/E_\mathrm{q}$, $\phi= 0.02$ and $\Gamma_0 = 0.0025/E_\mathrm{q}$.}
\label{fig:ergotropy}
\end{figure}

Using Eq.~\eqref{eq:ergotropy-final}, we can redefine the regimes of operation of the device reported in Sec.~\ref{sec:regimes} by replacing free-energy changes in the tape qubits, $\dot{F}_\mathrm{tape}$, by the corresponding ergotropy changes, $\dot{\mathcal{W}}_\mathrm{tape}$. In Fig.~\ref{fig:ergotropy}(a), we show the regions of the Bloch sphere corresponding to such new operational regimes for same parameters as in Fig.~\ref{fig:regimesI} of Sec.~\ref{sec:regimes}. By comparing the two figures, we notice that, as expected, the refrigerator regime (characterized by $\dot{Q}_\mathrm{c} > 0$) appears in the same region, while we observe differences for the shape of HE ($\dot{\mathcal{W}}>0$) and D ($\dot{\mathcal{W}}<0$) regions. In particular the HE regime broadens, including the whole south hemisphere. This indicates the existence of situations where the ergotropy of the tape qubits increases while their free energy decreases. While the opposite situation is not observed, there are, nevertheless, cases for which the free energy increases more than the ergotropy. For example, along the line $c=0$, the ergotropy changes vanishes in the south hemisphere while $\dot{F}_\mathrm{tape}$ is still positive.

In Fig.~~\ref{fig:ergotropy}(b) we plot the values of $\dot{\mathcal{W}}_\mathrm{tape}$ in units of $r \phi^2 E_\mathrm{q}$. Similarly to the case of free energy changes, we observe that highest values are reached for non-zero initial coherence of the tape qubits. More precisely, the maximum value of $\dot{\mathcal{W}}_\mathrm{tape}$ is reached for $\vert c \vert\sim 0.4$ and $p \simeq $, being $3$ times greater than the best incoherent case (reached at $p_1 = $ and $c=0$). Therefore, allowing initial coherence in the tape qubits enhances the output ergotropy current when maintaining all other resources constant. This confirms that input coherence plays a crucial role for the optimization of the performance of the machine, also from the point of view of ergotropy generation.

%%%%%%%%%%%%%%%%%%%%%%%%%%%%%%%%%%%%%%%%%%%%
\bibliographystyle{apsrev4-2-titles}
\bibliography{REFS.bib}

\begin{thebibliography}{94}
\bibitem [{not(pace)}]{note:levels}%
  \BibitemOpen
  \href@noop {} {} (\bibinfo {year} {{H}ere we denote $\ket{i j}_\mathrm{m}
  \equiv \ket{i}_\mathrm{c} \otimes \ket{j}_\mathrm{h}$ for $i,j= 0,1$ the
  energetic basis of the machine Hilbert space})\BibitemShut {NoStop}%
\bibitem [{not(work)}]{note1}%
  \BibitemOpen
  \href@noop {} {} (\bibinfo {year} {{F}or the definition of the free energy we
  employ the cold temperature $T_\mathrm{c}$ since it maximizes the extractable
  work})\BibitemShut {NoStop}%
\end{thebibliography}
%%%%%%%%%%%%%%%%%%%%%%%%%%%%%%%%%%%%%%%%%%%%

\end{document}